\documentclass[12pt,fleqn]{article}

\usepackage{epsfig}
\usepackage{amssymb}

\begin{document}

\begin{titlepage}


\begin{center}
{\Large \bf (In--)Consistencies in the relativistic description

\vspace{5mm}

of excited states in the Bethe--Salpeter equation}

\vspace{10mm}

        Steven Ahlig and Reinhard Alkofer \\
        Institut f\"ur Theoretische Physik,
        Universit\"at T\"{u}bingen, \\
        Auf der Morgenstelle 14, 72076 T\"ubingen, Germany

\end{center}

\vspace{10mm}
\begin{quote}
  {\bf Abstract}\\

The Bethe--Salpeter equation provides the most widely used technique to extract
bound states and resonances in a relativistic Quantum Field Theory. Nevertheless
a thorough discussion how to identify its solutions with physical states is
still missing. The occurrence of complex eigenvalues of the homogeneous
Bethe--Salpeter equation complicates this issue further. Using a perturbative
expansion in the mass difference of the constituents we demonstrate for scalar
fields bound by a scalar exchange that the underlying mechanism which results
in complex eigenvalues is the crossing of a normal (or abnormal) with an
abnormal state. Based on an investigation of the renormalization of
one--particle properties we argue that these crossings happen beyond the
applicability region of the ladder Bethe--Salpeter equation. The implications
for a fermion--antifermion bound state in QED are discussed, and a consistent
interpretation of the bound state spectrum of QED is proposed.

\end{quote}
\begin{quote}
  \end{quote}

\vfill
\end{titlepage}

\baselineskip=16pt plus 1pt minus 2pt

\section{Introduction}

In a relativistic Quantum Field Theory bound states and resonances are identified
through the occurrence of pole or cut like singularities in the Green's
functions of the theory. Thus it is natural to study the four--point Green's
function when searching for two--particle bound states. Indeed, the
corresponding equation has been proposed by Bethe and Salpeter \cite{BeSal51}
and subsequently been proven by Gell--Mann and Low \cite{GelLow51} and
Schwinger \cite{Schw51} as early as 1951. It can be interpreted as one of the
Dyson--Schwinger \cite{Schw51,Dys49} equations of Quantum Field Theory and it
is, thus, an inhomogeneous integral equation. Assuming the existence of a bound
state signaled by a pole in the four--point function the homogeneous
Bethe--Salpeter equation may be derived (see below). It allows to determine the
bound state masses and covariant wave functions. Usually one employs the
so-called ladder approximation which renders the homogeneous Bethe--Salpeter
equation in the form of an eigenvalue problem. Hereby the eigenvalue is
the square of the coupling constant, and the bound
state mass has to be tuned such that the Bethe--Salpeter eigenvalue equals the
given value of this constant. The covariant wave functions are then determined
as the eigenfunctions of the system.

The only analytically solvable example of a Bethe--Salpeter equation is the one
for two (massive) scalar particles bound by the ladder approximation to the
exchange of a massless scalar field \cite{Wick54,Cutk54,Naka69}. Despite its
relative simplicity as compared to realistic systems this model, the
Wick--Cutkosky model, displays already the advantages (e.g.\ full covariance)
as well as the shortcomings (e.g.\ the existence of abnormal states) inherent
to almost all Bethe--Salpeter based approaches used until today, for a recent
very accurate numerical solution of the ladder Bethe--Salpeter equation for
scalars see e.g.\ \cite{TjoNiew96} and references therein.

Common to the analytical and the numerical solutions is the existence of
abnormal states which have led to controversial discussions regarding
their physical interpretation \cite{Wick54,Cutk54,Naka69,Scarf,OhnWata}.
 Even worse, for the case
of constituents with unequal masses some eigenvalues of the homogeneous
Bethe--Salpeter equation become complex \cite{Kauf69,FukSeto93,FukSeto2}. Clearly, such
a behavior is unexpected and has to be understood. It is usually attributed to
the use of the ladder approximation which destroys crossing symmetry from the
very beginning. However, we will see in the following that already the use of
bare propagators is problematic. This will also not be cured by incorporating
one--particle self--energies and thereby generalising the ladder approximation.
Instead, it will be shown in this paper that the occurrence of complex
eigenvalues also happens with self--energies taken into account. It seems that
one has to start on a more fundamental level: One seriously has to raise the
question which values of the renormalized coupling constant are possible when
considering a {\it renormalized} Quantum Field Theory.

\subsection{Derivation of the homogeneous Bethe--Salpeter equation}
\label{Deri_hom_BSE}

In order to make this paper as self--contained as possible, and to make our
argumentation accessible to non--expert readers, we will supply some important
derivations and facts concerning the Bethe--Salpeter approach to relativistic
bound states in the remainder of this introduction. Readers familiar with the
Bethe--Salpeter equation probably will probably prefer to jump to Sec.\ 2 immediately.

\begin{figure}
\centerline{\epsfxsize 15.0cm\epsfbox{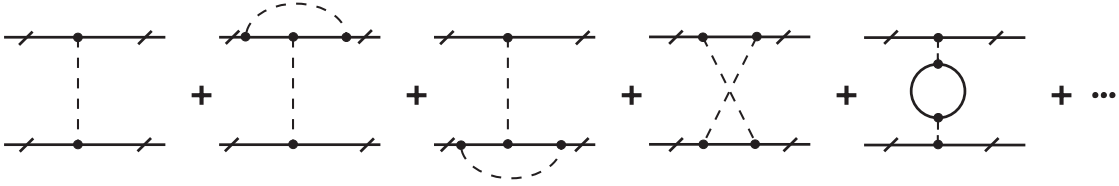}}
\caption{ Feynman diagrams of the first few terms in a
perturbative expansion of the kernel $K$. Solid lines represent propagators of
the constituents, dashed lines the propagator of the exchange particle.
\label{K1}}
\end{figure}

We assume to deal with three types of scalar fields. The two constituents are
supposed to have masses $m_1$ and $m_2$ and self--energies $\Sigma_1$ and
$\Sigma_2$. The four--point function $G^{\left( 4\right) }\left(
x_{1},x_{2},y_{1},y_{2}\right)$ describing the scattering of these two
constituents fulfills the inhomogeneous Bethe--Salpeter equation
\begin{equation}
\left[ \square _{x_{2}}+m_{2}^{2}-\Sigma _{2}\right] \left[ \square
_{x_{1}}+m_{1}^{2}-\Sigma _{1}\right] G^{\left( 4\right) }\left(
x_{1},x_{2},y_{1},y_{2}\right) =\quad \quad \quad \quad \quad \quad
\label{X1.16}
\end{equation}
\vspace{-0.4cm}
\begin{eqnarray*}
&&\quad \quad \quad \quad \delta ^{\left( 4\right) }\left( x_{1}-y_{1}\right)
\delta ^{\left( 4\right) }\left( x_{2}-y_{2}\right) +\delta ^{\left( 4\right)
}\left( x_{1}-y_{2}\right) \delta ^{\left( 4\right) }\left( y_{1}-x_{2}\right)
\\ &&\quad \quad \quad \quad \quad \quad \quad
\quad \quad \quad \quad +\int d^{4}z_{1}d^{4}z_{2}K\left(
x_{1},y_{1},z_{1},z_{2}\right) G^{\left( 4\right) }\left(
z_{1},z_{2},y_{1},y_{2}\right) .
\end{eqnarray*}
The kernel $K$ is defined as the sum of all amputated two--particle irreducible
contributions. The Feynman diagrams of the first few terms are depicted in fig.\
\ref{K1}.

For translationally invariant systems it is, of course, advantageous to
transform this equation to momentum space. Note that the introduction of the
relative coordinate $x=x_1-x_2$ allows an arbitrary parameter $\alpha \in [0,1]$
in defining the coordinate $X=\alpha x_1 + (1-\alpha )x_2$ which results in the
corresponding momenta
\begin{equation}
p_1=\alpha P +p \quad {\rm and} \quad p_2= (1-\alpha )P-p
\label{p12}
\end{equation}
for the constituents in terms of the total and relative momenta, $P$ and $p$,
respectively. Fourier transforming eq.\ (\ref{X1.16}) leads to
\begin{equation}
\int \frac{d^{4}p^{\prime }}{\left( 2\pi \right) ^{4}}\left[ {D}\left(
p,p^{\prime },P\right) +{K}\left( p,p^{\prime },P\right) \right] {G}^{\left(
4\right) }\left( p^{\prime },p^{\prime \prime },P\right)
=\delta ^{\left( 4\right) }\left( p-p^{\prime \prime }\right) ,
\label{inhBSE}
\end{equation}
where
\begin{equation}
{D}\left( p,p^{\prime },P\right) :=\left( 2\pi \right) ^{4}\delta
^{\left( 4\right) }\left( p-p^{\prime }\right) \left( {G}_{1}^{\left(
2\right) }\right) ^{-1}\left( p_{1}\right) \left( {G}_{2}^{\left( 2\right)
}\right) ^{-1}\left( -p_{2}\right) \label{X1.21}
\end{equation}
is defined in terms of the inverse two--point Green's functions of the
constituents.

\begin{figure}
\begin{center}
\begin{minipage}{20mm}
  \epsfig{file=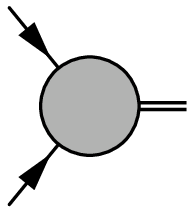,width=20mm}
\end{minipage}
\hspace{5mm}
=
\begin{minipage}{35mm}
  \epsfig{file=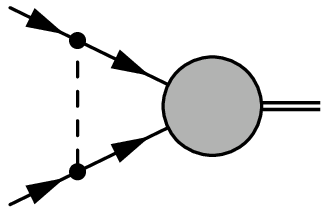,width=35mm}
\end{minipage}
\end{center}
\caption{Pictorial representation of the homogeneous Bethe--Salpeter equation in
         ladder approximation.
\label{BSE1}}
\end{figure}

The crucial step from the inhomogeneous to the homogeneous Bethe--Salpeter
equation consists in assuming a bound state reflecting itself in a pole in the
four--point Green's function for an on--shell momentum $P_{os}$, with
$P_{os}^{2} = M^2$
\begin{equation}
{G}^{\left( 4\right) }\left( p,p^{\prime },P_{os}\right)
=\frac{-i}{
\left( 2\pi \right) ^{4}}\frac{{\chi }\left( p,P_{os}\right)
{\bar{\chi}}\left( p^{\prime },P_{os}\right) }{2\omega \left( P^{0}-\omega
+i\epsilon \right) }+\rm{reg. \,terms},\,
\omega := \sqrt{ {\bf P}^{2} + M^2 }\,
\label{4point-pole}
\end{equation}
where we introduced the definition of the Bethe--Salpeter amplitudes
\begin{eqnarray}
\chi \left( x_{1},x_{2},P\right) &:=& \left\langle 0\left| T\left\{ \Phi
\left( x_{1}\right) \Phi \left( x_{2}\right) \right\} \right| P\right\rangle
\nonumber \\
\bar{\chi}\left( x_{1},x_{2},P\right) &:=& \left\langle 0\left| T \left\{ \Phi
^{\dagger }\left( x_{1}\right) \Phi ^{\dagger }\left( x_{2}\right) \right\}
\right| P\right\rangle \label{X1.30}
\end{eqnarray}
together with their Fourier transforms
\begin{eqnarray}
\chi \left( x_{1},x_{2},P\right) &=:&e^{-iPX}\int \frac{d^{4}p}{\left( 2\pi
\right) ^{4}}e^{-ipx}{\chi }\left( p,P\right)  \nonumber \\
\bar{\chi}\left( x_{1},x_{2},P\right) &=:&e^{+iPX}\int \frac{d^{4}p}{\left(
2\pi \right) ^{4}}e^{-ipx}{\bar{\chi}}\left( p,P\right).
\label{X1.31}
\end{eqnarray}
Hereby $\left| 0\right\rangle$ denotes the ground state (vacuum) and $\left|
P\right\rangle$ the bound state. Very close to the pole the regular terms can be
safely neglected and the dependence of the four--point function on the relative
momenta $p$ and $p^\prime$ can be separated. Expanding ${G}^{\left(4\right) }$
and $\left[ {D}-{K}\right] $ in the inhomogeneous Bethe--Salpeter equation in
powers of $\left( P^{0}-\omega \right) $ yields the homogeneous Bethe--Salpeter
equation and the normalisation condition for the amplitude. The order $\left(
P^{0}-\omega \right) ^{-1}$ provides
\begin{equation}
\int \frac{d^{4}p^{\prime }}{\left( 2\pi \right) ^{4}}\left[ {D}\left(
p,p^{\prime },P_{os}\right) +{K}\left( p,p^{\prime },P_{os}\right)
\right] {\chi }\left( p,P_{os}\right) =0,  \label{X1.35}
\end{equation}
whereas to ${\cal O}\left( \left(P^{0}-\omega \right)
^{0}\right) $ one obtains
\begin{eqnarray}
\int \frac{d^{4}p}{\left( 2\pi \right) ^{4}}\frac{d^{4}p^{\prime }}{\left(
2\pi \right) ^{4}}{\rm tr} \left({\bar{\chi}}\left( p,P_{os}\right)
\left. \frac{\partial }{\partial P^{0}}\left( D\left( p,p^{\prime },P\right)
+ K\left( p,p^{\prime },P\right) \right) \right| _{P^{0}=\omega }
\!\!\! {
\chi }\left( p^{\prime },P_{os}\right) \right) \nonumber \\
 = 2i\omega .
\label{X1.39}
\end{eqnarray}
This ensures the residue to be equal to $1$ at the bound state pole.

The homogeneous Bethe--Salpeter equation (\ref{X1.35}) is a linear integral
equation for the amplitude ${\chi}$ whose overall normalisation is fixed by
(\ref{X1.39}). Approximating the kernel by the one--boson--exchange depicted in
the first diagram of fig.\ \ref{K1} eq.\ (\ref{X1.35}) can be cast into an
eigenvalue problem for the coupling constant by defining the vertex function
$\Gamma
\left( p,P\right)$:
\begin{equation}
{\chi}\left( p,P\right) =:G_{1}\left( p_{1}\right) G_{2}\left( p_{2}\right)
\Gamma \left( p,P\right) . \label{X1.37}
\end{equation}
The homogeneous Bethe--Salpeter equation (\ref{X1.35}) then reads
\begin{equation}
\Gamma \left( p,P_{os}\right) = - \int \frac{d^{4}p^{\prime }}{\left( 2\pi
\right) ^{4}}K\left( p,p^{\prime },P_{os}\right) G_{1}\left( p_{1}^{\prime
}\right) G_{2}\left( p_{2}^{\prime }\right) \Gamma \left( p^{\prime
},P_{os}\right) .
\label{BSEvertex}
\end{equation}
In the ladder approximation the kernel $K$ is set equal to
\begin{equation}
  K(p,p') = i \frac{g^2}{(p-p')^{2} - \mu^{2}}
\end{equation}
which is just the propagator of the scalar exchange particle of mass $\mu$
multiplied with $g^{2}$. On inspection one finds that (\ref{BSEvertex})
is an eigenvalue problem for $g^{2}$ if $G_{1}$ and $G_{2}$ are the
bare propagators of the constituents.
The ladder approximation to the Bethe-Salpeter
equation is pictorially represented in fig. \ref{BSE1}.
 If a parameter pair $(g^2, P^0=M)$
exists the pole assumption is {\it a posteriori} justified and $M$ is the bound
state mass with $\chi$ being the corresponding amplitude (wave function) as can
be inferred from eq.\ (\ref{4point-pole}) which reflects, of course, nothing
else than the Lehmann representation of the four--point function.

\subsection{Abnormal solutions and relative time parity}
\label{sec. sym.}

\begin{figure}
\centerline{\epsfxsize 9cm\epsfbox{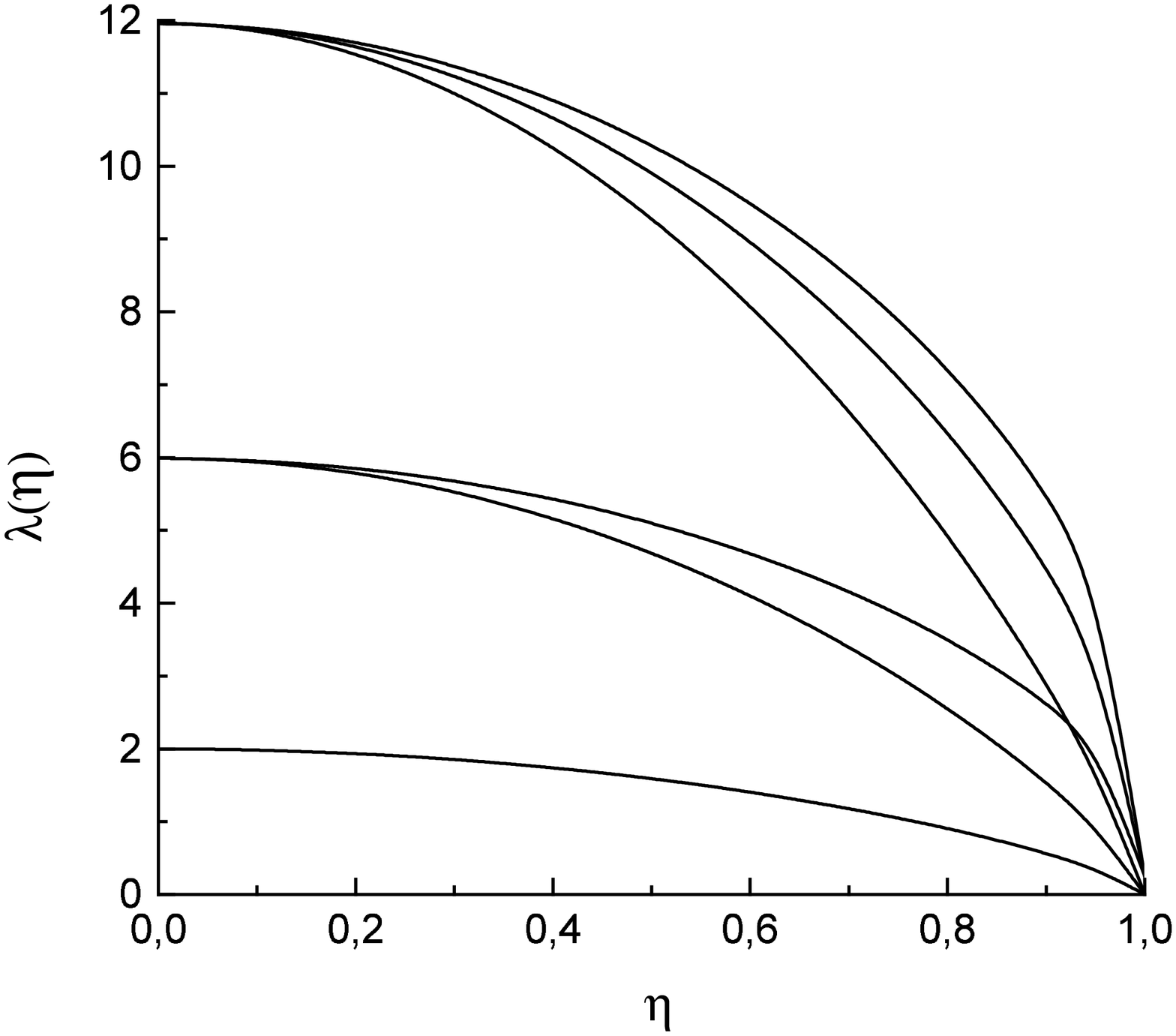}}
\caption{ The eigenvalues of the Bethe--Salpeter equation in the Wick--Cutksoky
model with constituents of equal masses as a function of the ratio of bound
state mass to the sum of masses of the constituents,$\eta
= M/2m$ .
\label{31}}
\end{figure}

\begin{figure}
\centerline{\epsfxsize 9cm\epsfbox{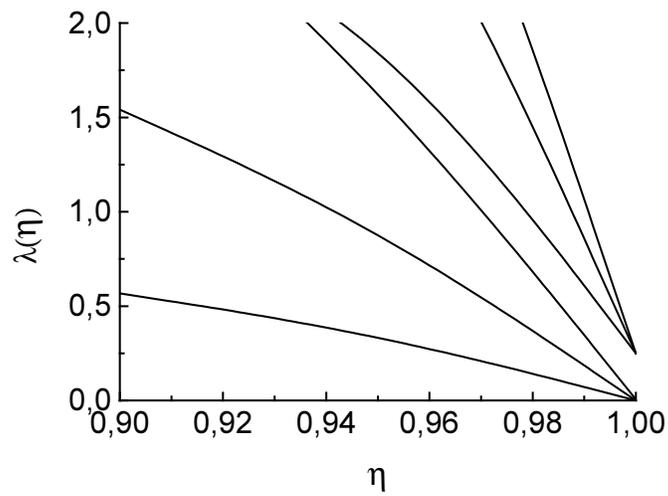}}
\caption{ Same as fig. \protect{\ref{31}}, however, for $\eta$ close to one.
\label{32}}
\end{figure}

For a given bound state mass $M$ the eigenvalue spectrum of the Bethe--Salpeter
equation should be positive definite, i.e.\ the eigenvalues should be real and
positive. We will see that this is not the case. Furthermore, one expects that
for very small coupling constants the binding energy vanishes and the bound
state mass becomes identical to the sum of the masses of the constituents. There
are states, however, which possess vanishing binding energy for a finite coupling.
They are therefore called
abnormal states. In the Wick--Cutksoky model \cite{ Wick54,Cutk54} with constituents
of equal masses
$m_1=m_2=m$ these abnormal states are easily identified: They only exist for
$\lambda :=g^{2}/16\pi ^{2} m^2
>\lambda _{c}=1/4$. If the binding energy becomes very small,
i.e.\ $\eta := M/2m \to 1$, the corresponding coupling constant $\lambda$
vanishes for the normal solutions, i.e.\ $\lambda \to 0$, whereas $\lambda \to
\lambda _{c}=1/4$ for the abnormal states, see fig.\ \ref{31} and its closeup
fig.\ \ref{32}.

On the other hand, in the opposite limit of a massless bound state, i.e.\ for
$\eta \to 0$, the eigenvalues are given by \cite{Cutk54}
\begin{equation}
\lambda =\left( n+\kappa \right) \left( n+\kappa +1\right) =N\left(
N+1\right) . \label{X2.15}
\end{equation}
The value of $\kappa $ equals the number of nodes of the eigenfunctions
when plotted as a function of the relative time $x_0$. For all normal solutions,
including the ground state, one has $\kappa =0$. This especially means that all
normal solutions are even under the reversal of relative time $x_{0}\rightarrow
-x_{0}$. On the other hand, $\kappa = 1,2, \ldots$ leads to
abnormal solutions, i.e.\ these solutions may be even or odd with respect
$x_{0}\rightarrow -x_{0}$:
\begin{equation}
\chi \left( -x_{0},\mathbf{x}\right) =
\left( -1\right) ^{\kappa }\chi \left(x_{0},\mathbf{x}\right) .
\label{X2.19}
\end{equation}
One may interpret the abnormal solutions as excitations in relative time. They
will obviously not appear in a purely non--relativistic treatment where the
constituents are considered for equal times only. However, not all abnormal
solutions necessarily vanish in a three--dimensional reduction of the
Bethe--Salpeter equation \cite{Bijt97}. On the contrary, the spectrum of a
three--dimensionally reduced equation will contain remnants of these abnormal
states.

\begin{figure}
\centerline{\epsfxsize 12cm\epsfbox{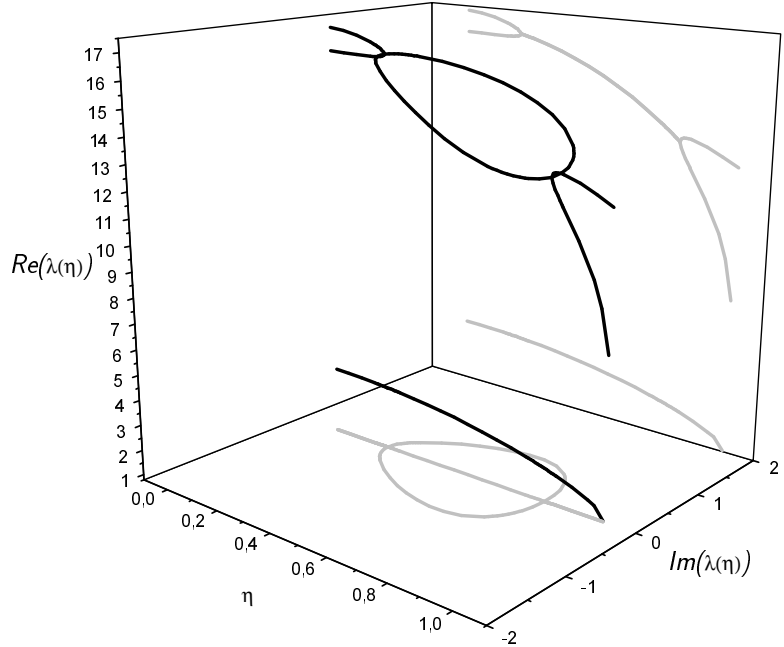}}
\caption{ The lowest lying three eigenvalues of the ladder Bethe--Salpeter
equation for $\delta=m_{1}/m_{2}=4$ and $\mu = m_2$.
\label{35}}
\end{figure}

For constituents of unequal masses and a massless exchange particle the
Bethe--Salpeter amplitudes are asymmetric, i.e.\ neither even nor odd, under the
transformation $x_{0}\rightarrow -x_{0}$. However, the eigenvalues of the
abnormal states start all at some common $\lambda_c$, see also \cite{Naka69} and
references therein.

For equally massive constituents and a massive exchange particle the amplitudes
do have a definite ``parity'' under the inversion of relative time. However,
there is no longer a common $\lambda_c$ at which the eigenvalues of the abnormal
solutions start.

The controversial discussion of the abnormal states already started with the
work of Wick \cite{Wick54} and Cutkosky \cite{Cutk54}, for some of the related
investigations see also \cite{Scarf,Ciaf,Naka69}. Fixing the spatial coordinates
of the constituents leaving only the relative time as degree of freedom one is
able to show that the abnormal solutions are all unphysical for this ``static''
model \cite{OhnWata}. However, it is not clear whether abnormal solutions can
in general be considered as unphysical \cite{Naka92}. As already mentioned, some of
these abnormal solutions survive most three--dimensional reductions
\cite{Bijt97}. Based on the analysis of several reductions of the
Bethe--Salpeter equation the main conclusion of ref.\ \cite{Bijt97} is that
abnormal solutions are very probably spurious consequences of the ladder
approximation supporting hereby an old conjecture by Wick \cite{Wick54}.

\begin{table}[tbp]
\begin{center}
\begin{tabular}{|c|c|c|}
\hline\hline
$~~~~~~~~~~~~~~M ,\mu ~~~~~~~~~~~~~~$ & symmetry group & Bethe--Salpeter
amplitude \\ \hline\hline $M =0,\mu =0$ & $O\left( 5\right) $ & $\chi \propto
Z_{nklm}\left(
\Omega _{5}\right) $ \\ \hline
$M =0,\mu \neq 0$ & $O\left( 4\right) $ & $\chi \propto Z_{klm}\left(
\Omega _{4}\right) $ \\ \hline
$M \neq 0,\mu =0$ & $O\left( 4\right) $ & $\chi \propto Z_{klm}\left(
\Omega _{4}\right) $ \\ \hline
$M \neq 0,\mu \neq 0$ & $O\left( 3\right) $ & $\chi \propto Y_{lm}\left(
\Omega _{3}\right) $ \\ \hline $\mu \rightarrow \infty $ & $O\left( 4\right) $ &
$\chi \propto Z_{klm}\left( \Omega
_{4}\right) $ \\ \hline
\end{tabular}
\end{center}
\caption{Summary of the symmetries of the scalar Bethe--Salpeter equation in
ladder approximation. $\mu$ denotes the mass of the exchange particle and $M$
is the Mass of the bound state. The functions $Z$ (or $Y$) denote the
spherical harmonics for the
corresponding $n$-sphere $\Omega_n$.
\label{t1}}
\end{table}

\subsection{Complex eigenvalues}

In this subsection we will consider the general case for a Bethe-Salpeter
equation containing only scalars: the masses of the constituents are assumed to
be unequal, the exchange particle is chosen to be massive. The numerical method
is detailed in appendix \ref{App. BSE1} and we will only give results here. For
comparable sets of parameters our results are equal to those given in refs.\
\cite{FukSeto93,FukSeto2}. In fig.\ \ref{35} the lowest three
eigenvalues for $\delta=m_{1}/m_{2}=4$ and $\mu = m_2$ are shown as a function
of $\eta = M/(m_1+m_2)$. The eigenvalue of the ground state is real for all
physically allowed values $\eta \in [0,1]$ and vanishes for $\eta\to 1$. The
ground state is thus a normal state. The situation differs drastically for the
higher--lying two states. At $\eta =1 $ the eigenvalues are non--vanishing, real
and positive. As the binding energy increases these two levels become degenerate
at $\eta \approx 0.81$. In the interval given approximately by $0.25 < \eta <
0.81$ the eigenvalues are complex with a degenerate real part and imaginary
parts of opposite sign. As one can see in fig.\ \ref{35} from the projection of
the curves not all eigenvalues are monotonically decreasing as a function of
$\eta$. Clearly, such a behaviour is unphysical because a decreasing coupling
constant should result in less binding.

\begin{figure}
\centerline{\epsfxsize 10cm\epsfbox{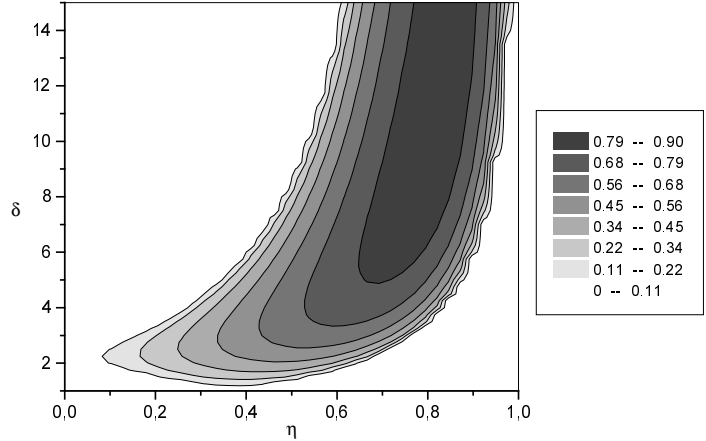}}
\caption{ The  absolute value of the imaginary part of the eigenvalues for the
first two excited states.
\label{36}}
\end{figure}

The appearance of complex eigenvalues is not restricted to a special
choice of parameters. In fig.\ \ref{36}
the absolute value of the imaginary part of the eigenvalues for the first two
excited states is shown for $\mu=m_2$ and $1< \delta \le 15$. One clearly sees
that with the increase in the mass of one constituent the interval in $\eta$ for
which the eigenvalues are non--real becomes smaller, however, the maximum value
of the imaginary part even increases. For a given mass ratio $\delta$ complex
eigenvalues exist for masses of the exchange particle up to some $\mu_{\rm
max}$, i.e.\ for $0< \mu \le \mu_{\rm max}$. Considering the first two excited
states we estimate this $\mu_{\rm max}$ to be
$\mu_{\rm max} \approx 0.85 m_1 + 1.66 m_2$.
Increasing the mass of the exchange particle the higher lying eigenvalues tend
to become real again. This can be understood from the fact that for an
infinitely heavy exchange particle the Bethe--Salpeter equation assumes an
$O(4)$ symmetric form as in the case of an massless exchange particle, see also
table \ref{t1} which summarises the symmetries of the scalar ladder
Bethe--Salpeter equation.

\goodbreak

\section{Inconsistency of the abnormal states}

In this section we relate the appearance of complex eigenvalues of the
Bethe--Salpeter-- (BS) equation to the presence of the abnormal solutions. To
this end we verify a conjecture of Kaufmann \cite{Kauf69}.

In section \ref{Deri_hom_BSE} we recalled the steps leading from the
inhomogeneous BS equation (\ref{X1.16}) to the homogeneous BS equation
(\ref{X1.35}). We now Wick-rotate the contour of integration in (\ref{X1.35})
according to $p_{0} \rightarrow ip_{4}, \mathbf{p}
\rightarrow \mathbf{p}$ and thus obtain the homogeneous BS equation
for euclidean momenta. A subsequent Fourier--transformation yields the
homogeneous BS equation in euclidean coordinate--space:
\begin{equation}
  {\cal L}\chi  = \left[ {\cal L}_{0}+{\cal P}\right] \chi  =
  \bar{\lambda}{\cal V}\chi .
  \label{X3.14}
\end{equation}
The operator ${\cal L}$ is the product of the inverse propagators of the
constituents and $\chi$ is the BS amplitude describing a bound state of two
scalar particles with masses $m_{1}$ and $m_{2}$, respectively. The interaction
of the two constituents is encoded in ${\cal V}$ which in the
ladder--approximation is
simply the Fourier--transform of the bare propagator of the exchange--particle.
We decomposed the operator ${\cal L}$ into two parts: ${\cal L}_{0}$,
containing the terms that are present for $m_{1}=m_{2}$ and ${\cal P}$
containing the additional terms that appear only for $m_{1} \neq m_{2}$. The
eigenvalue is denoted by $\bar{\lambda}$ instead of $\lambda$ for later
convenience. The explicit expressions are given by:
\begin{eqnarray}
  {\cal L}_{0} &=& \left[ \tilde{p}^{2}+2\eta \left( i\tilde{p}_{4}\right) +1-\eta ^{2}\right]
  \left[ \tilde{p}^{2}-2\eta \left( i\tilde{p}_{4}\right) +1-\eta ^{2}\right] \quad \quad \\
  &{\rm with }& \quad \tilde{p}_{\mu}=i\partial_{\mu}/m_{2} \nonumber \\
  \nonumber \\
  {\cal P} &=& \left( i\tilde{p}_{4}\right)
         \left[ 4\eta \Delta \left[ \tilde{p}^{2}-\left( 1-\eta ^{2}\right) \right] +
         \Delta
         ^{3}4\eta \left( 1-\eta ^{2}\right) \right] \nonumber \\
         &+&\Delta ^{2}\left[ 2\left( 1-\eta ^{2}\right)
         \left[ \tilde{p}^{2}-\left( 1-\eta ^{2}\right) \right] -4\eta ^{2}\tilde{p}_{4}^{2}
         \right]
         +\Delta ^{4}\left( 1-\eta ^{2}\right) ^{2} \label{Pexpl}\\
  \nonumber \\
  {\cal V} &=& \frac{1}{\pi ^{2}}\int d^{4}q\frac{e^{-iq\cdot x}}{%
  q^{2}+\mu ^{2}} \\
  \bar{\lambda} &=& \frac{g^{2}}{16\pi^{2} m_{2}^{2}}
  \label{X3.4}
\end{eqnarray}
The parameter $\eta= M/(m_{1}+m_{2})$ measures the mass of the bound state,
$\Delta=(m_{1}-m_{2})/2m_{2}$ is proportional to the mass difference, and
$\mu$ is the mass of the exchange--particle.

The explicit expression (\ref{Pexpl}) shows that ${\cal P}$ vanishes for a
vanishing mass difference $\Delta$. Thus for small $\Delta$ the decomposition ${\cal
L}={\cal L}_{0}+{\cal P}$ can be understood as dividing ${\cal L}$ into the
dominant part ${\cal L}_{0}$ and into the perturbation ${\cal P}$. In the
following we will formalise this point of view and investigate the consequences
thereof.

Suppose $\chi _{1}$ and $\chi _{2}$ are two solutions of the homogeneous BS
equation for constituents of equal mass with eigenvalues $\lambda _{1}$ and
$\lambda _{2}$ respectively:
\begin{eqnarray}
{\cal L}_{0}\chi _{1} &= \lambda _{1}{\cal V}\chi _{1} \nonumber \\
{\cal L}_{0}\chi _{2} &= \lambda _{2}{\cal V}\chi _{2}  \label{eq210}
\end{eqnarray}
Because of the restriction $m_{1}=m_{2}$ the BS amplitudes $\chi
_{1}$ and $\chi _{2}$ have a definite $x_{0}$-parity. As we will
see the interesting case is to take $\chi _{1}$ and $\chi
_{2}$ to be of opposite $x_{0}$--parity, and this is assumed for the following
considerations.
Since we are interested in the region where two real solutions become a pair
of complex conjugated solutions we take the ansatz $\chi=a\chi _{1}+b\chi _{2}$
where $\chi$ is the BS amplitude for different masses $m_{1} \neq m_{2}$ and
$\chi _{1},\chi _{2}$ are the amplitudes for $m_{1}=m_{2}$ as defined in
(\ref{eq210}). Substituting this ansatz for $\chi$ into (\ref{X3.14}) and
projecting onto $\chi _{1}$ and $\chi _{2}$ one obtains a system of equations
relating the eigenvalue $\bar{\lambda}$ to the eigenvalues
$\lambda_{1},\lambda_{2}$:
\begin{eqnarray}
a\lambda _{1}{\cal V}_{11}+a{\cal P}_{11}+b{\cal P}_{12}
 &= \bar{\lambda}a{\cal V}_{11} \nonumber \\
b\lambda _{2}{\cal V}_{22}+a{\cal P}_{21}+b{\cal P}_{22}
 &= \bar{\lambda}b{\cal V}_{22} \label{210}
\end{eqnarray}
where
\begin{equation}
  X_{ik}=\langle \chi_{i}, X \chi_{k} \rangle :=
  \int d^{4}x\,\chi_{i}^{*}\left(
  x\right) X \chi_{k} \left( x\right)
\end{equation}
for $X={\cal P},{\cal V}$ and $i,k=1,2$. In deriving (\ref{210}) we used ${\cal
V}_{12}= {\cal V}_{21}=0$ which is due to the different $x_{0}$--parity of
$\chi_{1}$ and $\chi _{2}$.

Assuming $\chi _{1}$ and $\chi _{2}$ to be
normalised
according to ${\cal V}_{11}={\cal V}_{22}=1$ one finds the two eigenvalues
$\bar{\lambda}_{1,2}$ of the BS equation (\ref{X3.14}),
\begin{equation}
  \bar{\lambda}_{1,2}=\frac{1}{2}\left( \lambda _{1}+\lambda
  _{2}+{\cal P}_{11}+{\cal P}_{22}\right) \pm \sqrt{\frac{1}{4}
  \left( \lambda _{1}-\lambda
  _{2}+{\cal P}_{11}-{\cal P}_{22}\right) ^{2}+{\cal P}_{12}{\cal P}_{21}} \,\,.
  \label{eq: neue Ew}
\end{equation}
\begin{figure}[t]
\centerline{\epsfxsize 10cm\epsfbox{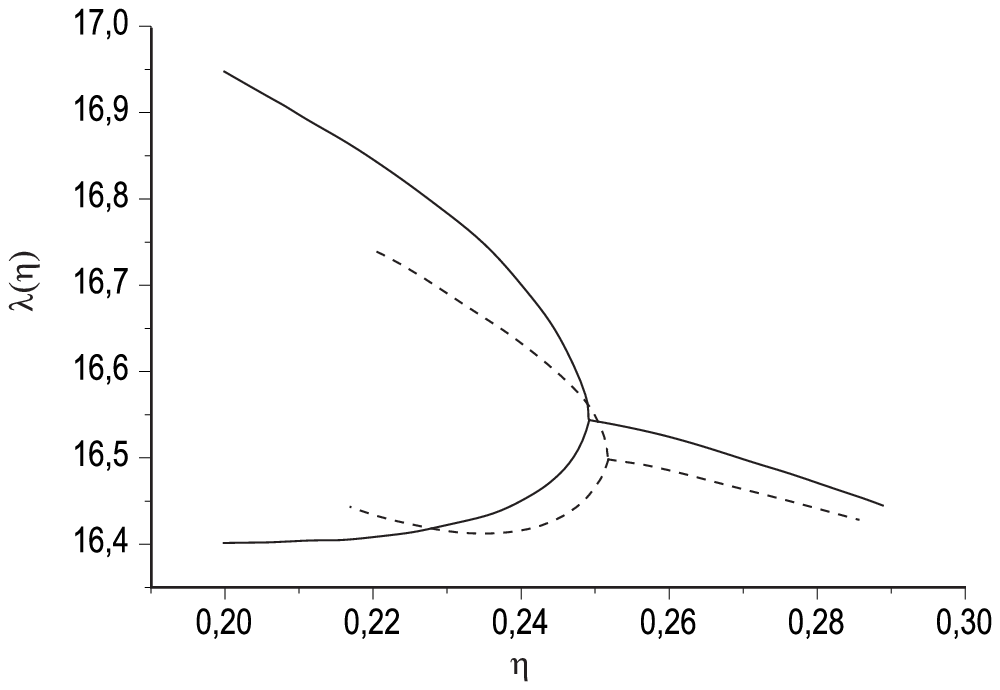}}
\caption{ The real part of two colliding solutions, according to the
numerical solution (solid line) and its semi--analytic estimate
\protect{(\ref{eq: neue Ew})} (dashed line), c.f.\ also the upper
left corner in fig.\
\protect{\ref{35}}.
\label{Abb41}}
\end{figure}
This gives $\bar{\lambda}$, i.e.\ the eigenvalue for $\Delta \neq 0$, in terms of
the matrix-elements ${\cal P}_{ik}$ and $\lambda_{1},\lambda_{2}$ which are the
eigenvalues for $\Delta = 0$. Using the explicit expression (\ref{Pexpl}) one
may calculate the matrix elements ${\cal P}_{ik}$ thereby employing the
solutions $\chi _{1}$ and $\chi _{2}$ obtained by numerically solving the BS
equation for $\Delta = 0$. On inspection one finds that ${\cal P}$ is
antihermitian leading to ${\cal P}_{12}=-{\cal P}_{21}^{*}\Rightarrow {\cal
P}_{12}{\cal P}_{21}=-
\left| {\cal P}_{12}\right| ^{2}$. According to (\ref{eq: neue Ew}) this
is a necessary condition for the appearance of complex eigenvalues. This is in
contrast to the hermitian perturbations that are considered usually. Whereas a
hermitian perturbation will enforce a repulsion between the solutions (avoided
crossing) an antihermitian perturbation will result in an attraction.

The real part of $\bar{\lambda}_{1,2}$ as calculated according to (\ref{eq: neue
Ew}) together with the result of the numerical calculation is shown in fig.
\ref{Abb41}. The dashed line represents the results according to the
coupling mechanism that has been proposed in this section whereas the solid line
represents the results of the numerical solution of the BS equation with $\delta
=m_1/m_2=4$ and $\mu=m_2$, c.f.\ the upper left corner in fig.\ \ref{35}. The two
curves are in reasonable quantitative agreement. The simple ansatz $\chi=a\chi
_{1}+b\chi _{2}$ and the coupling due to ${\cal P}$ reproduce the
overall appearance of the collision astonishingly well. The imaginary parts of
the solutions $\bar{\lambda}(\eta)$ appear after the collision point. They
agree with those of the numerical solution to the same level as the real parts.
Therefore we conclude: the coupling of solutions of opposite $x_{0}$-parity
\emph{is} the mechanism leading to complex eigenvalues. Hereby,
the presence of solutions of positive \emph{and} negative $x_{0}$-parity is a
prerequisite for the appearance of complex eigenvalues $\bar{\lambda}$.
Choosing both states to be of the same $x_0$--parity will lead to an avoided
crossing. Thus, at least one abnormal state is necessary for the occurence
of a pair of complex eigenvalues.

In this context it is interesting to note that
Naito and Nakanishi \cite{NaitoNaka} derived a normalization condition
which they claimed to remove the solutions with complex eigenvalues.
However, a substantial loophole has been spotted in their derivation
\cite{Ida} and Fukui and Set\^{o} \cite{FukSeto2} showed numerically that the
normalization condition can be satisfied for solutions with real as
well as for solutions with complex eigenvalues.

In the
following we will show that the solutions with negative $x_{0}$-parity lead to
inconsistencies on a even more fundamental level. We will derive an equation
that has to be satisfied by all solutions of the BS equation in
ladder approximation which, however, is only satisfied by the solutions with
positive $x_{0}$--parity, see also refs.\ \cite{Nak23,CtkLeo11}.

The equation may be derived starting with the inhomogeneous BS equation
\begin{equation}
  \bar{D}\left( p,P\right) G\left( p,q,P\right) =\delta ^{4}\left( p-q\right)
  +\int
  d^{4}p^{^{\prime }}\,K\left( p,p^{\prime },P\right) G\left( p^{\prime
  },q,P\right)
  \label{eq: Norm 1}
\end{equation}
where $\bar{D}$ is the product of inverse propagators
\begin{equation}
  \bar{D}\left( p,P\right) := G_{0}^{-1}\left(p_{1}\right) G_{0}^{-1}\left(
  p_{2}\right) , \quad  p_{1} := \alpha P + p ,
  \quad p_{2} := (1-\alpha) P - p ,
  \label{X4.1}
\end{equation}
and is slightly different from $D$ defined in (\ref{X1.21}). The following
considerations are restricted to the case $m_{1}=m_{2}$ which is reflected in
the definition (\ref{X4.1}). $G$ is the 4-point function. It depends on
the total 4-momentum $P$ and the relative 4-momenta $p$ and $q$. The kernel $K$ is
defined as the sum over all 2--particle irreducible diagrams. Taking the
derivative with respect to $\lambda = g^{2}/16\pi^2 m_{2}^{2}$ one obtains
after some reordering
\begin{eqnarray}
\int d^{4}p\,d^{4}p^{\prime }\,G\left( p^{\prime \prime },p,P\right) \left[
\frac{\partial }{\partial \lambda }\bar{D}\left( p,P\right) \delta ^{4}\left(
p^{\prime }-p\right) -\frac{\partial }{\partial \lambda }K\left( p,p^{\prime
},P\right) \right] G\left( p^{\prime },q,P\right) \nonumber \\
=-\frac{\partial }{\partial \lambda }G\left( p^{\prime \prime },q,P\right)\,.\quad
\label{eq: Norm 3}
\end{eqnarray}
No approximation has been used so far. To proceed we restrict the
considerations to the
case of bare propagators and to the use of the ladder approximation for the
kernel $K$. This yields
\begin{eqnarray}
  \frac{\partial }{\partial \lambda }\bar{D}\left( p,P\right) &=& 0  ,
  \nonumber \\
  \frac{\partial }{\partial \lambda }K\left( p,p^{\prime },P\right) &=& \frac{1}
  {\lambda }K\left( p,p^{\prime },P\right) .
  \label{X4.7}
\end{eqnarray}
Substituting the representation of the 4-point function as given
in eq.\ (\ref{4point-pole}) and comparing the pole contributions one obtains
\begin{equation}
  i\int d^{4}p\,d^{4}p^{\prime }\,\overline{\chi }\left( p\right) K\left(
  p,p^{\prime },P\right) \chi \left( p^{\prime }\right) =\lambda \frac{dM^{2}}
  {d\lambda }
  \label{eq: Norm4}
\end{equation}
where $\chi$ is the BS amplitude. Equation (\ref{eq: Norm4}) should be satisfied
by all solutions $\chi$ of the BS equation if bare propagators and the
ladder approximation for the kernel $K$ are used.

One can make use of translation invariance to factor out the center of mass
motion:
\begin{equation}
  \chi = {\rm e}^{-i P \cdot X} \phi \left( x,P \right).
\end{equation}
The coordinates are hereby defined according to
\begin{eqnarray}
  x := x_{1} - x_{2}, \quad  X := \alpha x_{1} + (1-\alpha)x_{2}
\end{eqnarray}
and the $x_{0}$--parity is reflected in the relation $\phi \left(
-x^{0},\mathbf{x}\right) =\epsilon \left( \phi \right) \,\phi
\left( x^{0},\mathbf{x}\right)$ with $\epsilon \left( \phi \right) =\pm 1$.
In the next step one decomposes the amplitudes $\phi$ into positive and negative
frequency parts, see \cite{Wick54}. After Wick rotation,
\begin{equation}
  p:=(p_{0},{\bf p}) \rightarrow (ip_{4},{\bf p})=:\bar{p}, \qquad
  (\bar{p}^{2}>0) ,
  \label{Wick-rot}
\end{equation}
one obtains the relation, see appendix \ref{app. relat.} for its derivation,
\begin{equation}
  \bar{\phi} ( \bar{p},P ) = \epsilon (\phi) \left[ \phi(\bar{p},P) \right]^{*}
\label{X4.24}
\end{equation}
for the Euclidean amplitudes. Applying the Wick rotation (\ref{Wick-rot})
to (\ref{eq: Norm4}) and substituting the relation (\ref{X4.24}) one arrives at
\begin{equation}
-\epsilon \left( \phi \right) \int d^{4}\bar{p}\,d^{4}\bar{p}^{\prime
}\,\left[ \phi \left( \bar{p},P\right) \right]
^{*}K^{\prime
}\left( \bar{p},\bar{p}^{\prime },P\right) \phi \left( \bar{p}%
^{\prime },P\right) =\lambda \frac{dM^{2}}{d\lambda }
\label{X4.25}
\end{equation}
with positive $K^{\prime }\left( \bar{p},\bar{p}^{\prime },P\right)
=\left( \bar{p}-\bar{p}^{\prime }\right) ^{-2}$.
As was discussed in the introduction a decreasing coupling constant should
result in less binding, i.e.\ $dM^{2}/d\lambda$ should be negative.
Eq.\ (\ref{X4.25}) implies that this is the case only for
$\epsilon \left( \phi \right) =+1$,
i.e.\ only the states with a positive $x_{0}$--parity can
satisfy both, eq.\  (\ref{X4.25}) and the physical condidtion
$dM^{2}/d\lambda <0$.

On the one hand, we have obtained eq.\ (\ref{eq: Norm4}) that has been derived
from the exact BS equation using bare propagators and the ladder approximation
for the kernel. On the other hand, we found that the solutions of negative
$x_{0}$--parity (being solutions of the BS equation in the same approximation)
cannot satisfy this equation and behave simultaneously reasonable under the
change of the coupling constant. As a matter of fact, we found in our numerical
results examples for both types of behaviour, i.e.\ either a violation of  eq.\
(\ref{X4.25}) or a positive slope when ploting $M^2$ vs.\  $\lambda$.  As a
subset of the abnormal solutions the solutions of negative $x_{0}$--parity are
present only above a certain nonzero coupling-constant. To reconcile these
conflicting results one has to question  the combined use of free propagators
and the ladder approximation for coupling  constants that are above this
critical value.

\goodbreak

\section{Renormalization of one--particle properties}

In this section we first present a consistent approximation scheme for the
propagators and the kernel of the BS equation. The resulting system of
Dyson--Schwinger equations is solved yielding a domain of validity for the
ladder approximation. The equations are derived using the formalism of
Cornwall, Jackiw and Tomboulis (CJT) \cite{CJT}.

The starting point is the generating functional
\begin{equation}
  Z[J,K]:=\int \prod_{i} {\cal D}\Phi_{i} \exp\left( i\left[ S[\Phi]+\Phi\cdot J
          + \frac{1}{2} \Phi\cdot K \cdot \Phi \right] \right)
  \label{eq: CJT Def. Z}
\end{equation}
where $\Phi_{i}(x)$ are scalar fields, $S[\Phi]$ is the classical action and the
abbreviations are defined by
\begin{eqnarray}
  \Phi \cdot J
  &:=& \sum_{i} \int d^{4}x\,\Phi _{i}\left( x\right)
  J^{i}\left( x\right) ,    \nonumber \\
  \Phi \cdot K\cdot \Phi
  &:=& \sum_{i,k} \int d^{4}x\,d^{4}y\,\Phi _{i}\left( x\right)
  K^{ik}\left( x,y\right) \Phi _{k}\left( y\right) .
  \label{X5.1}
\end{eqnarray}
The generating functional for connected Greens functions is defined as usual by
$Z\left[ J,K\right] =:\exp \left( iW\left[ J,K\right]
\right)$. Using
\begin{eqnarray}
  \label{eq: CJT Def. G}
  \frac{\delta W\left[ J,K\right] }{\delta J^{i}\left( x\right) } &=:& \phi
  _{i}\left( x\right) \nonumber \\
  \frac{\delta W\left[ J,K\right] }{\delta K^{ik}\left( x,y\right) } &=:&
  \frac{1}{%
  2}\left( \phi _{i}\left( x\right) \phi _{k}\left( y\right) +G_{ik}\left(
  x,y\right) \right)
\end{eqnarray}
and performing a double Legendre transformation one obtains the effective action
$\Gamma$ as a functional of $\phi$ and $G$,
\begin{equation}
\Gamma \left[ \phi ,G\right] :=W\left[ J,K\right] -\phi \cdot J-\frac{1}{2}%
\phi \cdot K\cdot \phi -\frac{1}{2}G\cdot K .  \label{X5.9}
\end{equation}
Here the notation defined in (\ref{X5.1}) has been used.
In ref.\ \cite{CJT} the expression
\begin{equation}
\Gamma \left[ \phi ,G\right] =S\left[ \phi \right] +\frac{1}{2}i{\rm %
Tr\,Ln}\left( G^{-1}\right) +\frac{1}{2}i {\rm Tr}\left( {\cal D}%
^{-1}G\right) +\Gamma _{2}\left[ \phi ,G\right]
\label{eq: CJT Entw. Gamma}
\end{equation}
has been derived where ${\cal D}^{-1}$ is given by
\begin{equation}
i{\cal D}_{ik}^{-1}\left( x,y;\phi \right) := iD_{i}^{-1}\left( x-y\right)
\delta _{ik}+\frac{\delta ^{2}S_{int}\left[ \phi \right] }{\delta \phi
_{i}\left( x\right) \delta \phi _{k}\left( y\right) } , \label{X5.16}
\end{equation}
and $D$ is the bare propagator. The nontrivial part in the effective action
(\ref{eq: CJT Entw. Gamma}) is $\Gamma_{2}$. It is defined as a sum of
vacuum loops where the
vertices are given by the interaction part of $S[\Phi+\phi]$ and the
propagators are equal to $G_{ik}$.

The BS equation for the amplitude $\psi$ in its most general form is given by
\begin{equation}
\int d^{4}x^{\prime }\,d^{4}y^{\prime }\,\Gamma _{ik,i^{\prime }k^{\prime
}}^{\left( 2\right) }\left( x,y;x^{\prime },y^{\prime }\right) \psi
^{i^{\prime }k^{\prime }}\left( x^{\prime },y^{\prime }\right) =0
\label{eq: CJT allg. BSE 2}
\end{equation}
where $\Gamma _{ik,i^{\prime }k^{\prime}}^{\left( 2\right) }\left( x,y;x^{\prime
},y^{\prime }\right)$ is the 2nd functional derivative of $\Gamma$ with respect
to $G$. The propagators $G$ are the solution of the corresponding
Dyson-Schwinger equations
\begin{equation}
\frac{\delta \Gamma \left[ \phi ,G\right] }{\delta G_{ab}\left( x,y\right) }
=0 . \label{X5.20}
\end{equation}

\begin{figure}
\begin{center}
$\biggl(
\begin{minipage}{20mm}
  \epsfig{file=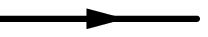,width=20mm}
\end{minipage}
\biggr)^{-1}
=\hspace{4mm}
\biggl(
\begin{minipage}{20mm}
  \epsfig{file=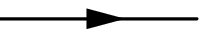,width=20mm}
\end{minipage}
\biggr)^{-1}
+\hspace{4mm}
\begin{minipage}{2cm}
  \epsfig{file=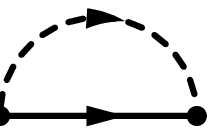,width=2cm}
\end{minipage}
$
\end{center}
\vspace{-2cm}
\begin{center}
$\biggl(
\begin{minipage}{20mm}
  \epsfig{file=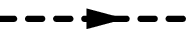,width=20mm}
\end{minipage}
\biggr)^{-1}
=\hspace{4mm}
\biggl(
\begin{minipage}{20mm}
  \epsfig{file=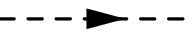,width=20mm}
\end{minipage}
\biggr)^{-1}
+\hspace{4mm}
\begin{minipage}{2.0cm}
  \hspace{0.2cm}
  \epsfig{file=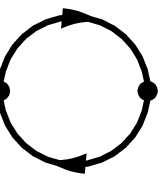,width=1.4cm}
\end{minipage}
$
\end{center}
\caption{ Graphical representation of the coupled system of Dyson-Schwinger
          equations
          (\ref{eq: CJT DSE expl.12P}), (\ref{eq: CJT DSE expl.3P}).
          Thick and thin lines represent the dressed and the bare propagators,
          respectively. }
\label{DSE_123_boson}
\end{figure}

In the following we will consider
a theory of three scalar fields $\Phi_{1},\Phi_{2}$ and $\Phi_{3}$ describing
massive particles and interacting according to the Lagrangian
\begin{equation}
 {\cal L} = \sum_i\frac{1}{2} \partial_{\mu} \Phi_{i}
 \partial^{\mu} \Phi_{i} - \sum_i \frac{1}{2} m_{i}^{2} \Phi_{i}^{2}
 +\frac{g}{\sqrt{2}} \left( \Phi_{1}^{2} +\Phi_{2}^{2}\right) \Phi_{3}.
 \label{Lscal}
\end{equation}
We approximate $\Gamma_{2}$ by the first nontrivial term of the loop-expansion,
i.e.\ we use
\begin{eqnarray}
\Gamma_{2} &=&
\begin{minipage}{12mm}
\epsfig{file=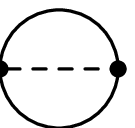,width=10mm}
\end{minipage}
\nonumber \\
 &=& \frac{1}{4}g^{2}\int d^{4}x^{\prime
 }\,d^{4}y^{\prime }\,\left( \sum_{i,k=1}^{2}G_{ik}^{2}(x',y') G_{33}(x',y') \right. \\
 &&\quad\quad\quad \quad\quad\quad \,\,\,\,\, \left. +\sum_{i=1}^{2}
 G_{ii}^{2}(x',y') G_{33}(x',y') \right) \, . \nonumber
\end{eqnarray}
This approximation for $\Gamma_{2}$ results in the rainbow--ladder approximation
for the BS equation (\ref{eq: CJT allg. BSE 2}) and the Dyson--Schwinger
equations (\ref{X5.20}). Having evaluated the functional derivatives we
Fourier transform the resulting system of equations to momen\-tum--space thus
obtaining the BS equation
\begin{equation}
  G_{1}^{-1}\left( p_{1}\right) G_{2}^{-1}\left(
  p_{2}\right) \psi^{12}\left( p,P\right) = g^{2}
  i\int \frac{d^{4}q}{\left( 2\pi \right) ^{4}}\,G_{3}\left(
  p-q\right) \psi^{12}\left( q,P\right)
  \label{eq: CJT BS expl. 1P}
\end{equation}
for the BS amplitude $\psi^{12}$ describing a bound states of particles 1 and
2. The momenta are defined by
\begin{eqnarray}
p_{1} := p-\alpha P, \quad  p_{2} := p+\left( 1-\alpha \right) P .
\end{eqnarray}
Evaluating explicitely the functional derivative (\ref{X5.20})
we obtain the coupled system of Dyson-Schwinger equations:
\begin{eqnarray}
  G_{i}^{-1}\left( p\right) &=& D_{i}^{-1}\left( p\right)
  -2g^{2}i\int \frac{d^{4}q}{\left( 2\pi \right) ^{4}}\,G_{3
  }\left( p-q\right) G_{i}\left( q\right) \quad \quad i=1,2
  \label{eq: CJT DSE expl.12P}\\
  G_{3}^{-1}\left( p\right) &=& D_{3}^{-1}\left(
  p\right) -g^{2}i\int \frac{d^{4}q}{\left( 2\pi \right) ^{4}}\,
  \sum_{i=1,2}
  G_{i}\left( p-q\right) G_{i}\left( q\right)
  \label{eq: CJT DSE expl.3P}
\end{eqnarray}
for the propagators $G_i$ that are used in the BS equation. This system of
equations is depicted diagrammatically in fig.\ (\ref{DSE_123_boson}).

Up to now we have used the unrenormalized fields $\Phi_{i}$, masses
$m_{i}$ and coupling constant $g$. Introducing the corresponding
renormalized quantities $\bar{\Phi}_{i} , \bar{m}_{i} , \bar{g}$ through
\begin{eqnarray}
  \sqrt{Z_{i}}\bar{\Phi}_{i} = \Phi _{i} , \quad
  Z_{m_{i}}\bar{m}_{i}^{2}   = m_{i}^{2} , \quad
  Z_{g}\bar{g}               = g
  \label{X5.28}
\end{eqnarray}
one is able to repeat the steps leading from (\ref{X5.20}) to (\ref{eq: CJT DSE
expl.12P}) and (\ref{eq: CJT DSE expl.3P}). As we use bare vertices (rainbow
approximation) we have to set $Z_g=1$ for consistency. This procedure
yields the Dyson-Schwinger equations for the renormalized self-energies
$\Sigma(p^{2}):=p^{2}-m^{2}-G^{-1}(p)$
\begin{eqnarray}
\Sigma _{i}\left( p^{2}\right) &=& \Delta_{i} +Z_{i}
2g^{2}i\int \frac{d^{4}q}{\left( 2\pi \right) ^{4}}\,G_{3}\left( p-q\right)
G_{i}\left( q\right) \quad \quad i=1,2
\label{eq: CJT DSren 1,2} \\
\Sigma _{3}\left( p^{2}\right) &=& \Delta_{3}  +%
g^{2}i\int \frac{d^{4}q}{\left( 2\pi \right) ^{4}}\,
\sum_{i=1,2} Z_{i}
G_{i}\left( p-q\right) G_{i}\left( q\right)
\label{eq: CJT DSren 3}
\end{eqnarray}
where the abbreviation
\begin{equation}
\Delta _{i}:=\left( 1-Z_{i}\right) p^{2}+\left(
Z_{i}Z_{m_{i}}-1\right) m_{i}^{2} \qquad i=1,2,3
  \label{X5.27}
\end{equation}
has been used. This coupled system of Dyson--Schwinger equations is finally
defined by  fixing the renormalization constants according to the on--shell
renormalization
\begin{equation}
\Sigma _{i}\left( m_{i}^{2}\right) =\left. \frac{d}{dp^{2}}\Sigma _{i}\left(
p^{2}\right) \right| _{p^{2}=m_{i}^{2}}=0 .\label{eq: CJT Renorm.bedg.}
\end{equation}
Note that the Lagrangian (\ref{Lscal}) defines a super--renormalizable model.
Thus, the mass renormalization constants  $Z_{m_{i}}$ are logarithmically
instead of quadratically divergent, the field  renormalization constants
$Z_{i}$ are finite. We take them nevertheless into account in order to mimic
an exactly renormalizable theory as closely as possible.

In order to obtain a numerical solution a Wick rotation to Euclidean space
is performed. The resulting coupled system of integral equations is then solved
iteratively. The details of the numerical method are given in appendix \ref{app. SS DSE}.

Since we are interested in the domain of validity of the
ladder approximation we focus on the physical constraints to be satisfied by the
propagators. The quanta of the fields $\Phi$ are assumed to describe physical
particles. This implies that the associated renormalization constants
$Z_i$ and $Z_{m_{i}}$ are positive and smaller than 1, i.e.\ $0 \leq Z_i,Z_{m_{i}} \leq 1$.
\begin{figure}
\epsfig{file=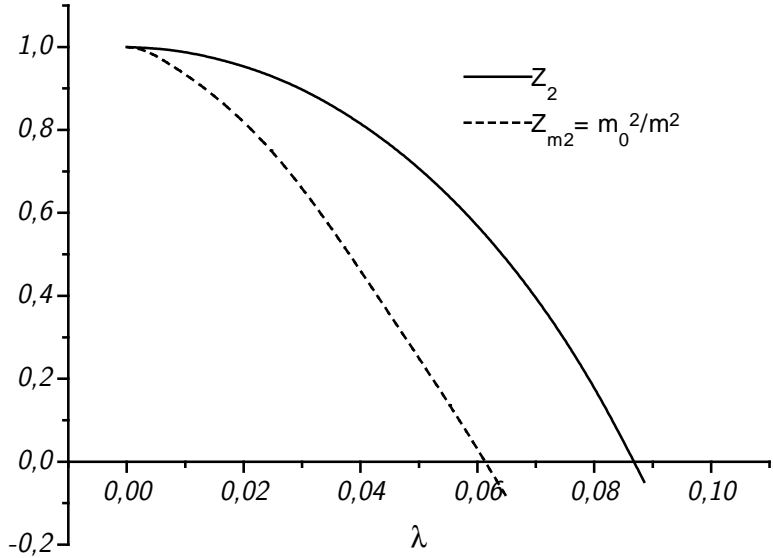,width=8cm, angle=-90}
\caption{ The dependence of the renormalization constants $Z_{2} ( \mu = m_{2} )$
and $Z_{m_{2}} (\Lambda_{UV} , \mu = m_{2} )$ on
$\lambda = g^{2}/16\pi^{2} m_{2}^{2}$. The renormalized masses
of the constituents were taken to be $m_{1}/m_{2}=4$ and the mass of the
exchange particle was set equal to $m_{2}$.}
\label{Abb63}
\end{figure}

The dependence of the renormalization constants $Z_{i}$ and $Z_{m_{i}}$
on the coupling parameter
$\lambda=g^{2}/16\pi^{2} m_{2}^{2}$ displays an interesting behaviour. Fig.
\ref{Abb63} shows the results for $Z_{2}$ and $Z_{m_{2}}$ which are the
renormalization
constants for  the less massive field $\Phi_{2}$. The masses of the constituents
were taken to be $m_{1}/m_{2}=4$ and the mass of the exchange particle was set
equal to $m_{2}$. $Z_{2}$ and $Z_{m_{2}}$ start at one for small values
of the coupling constant and decrease for greater values. Both, the mass
renormalization $Z_{m_{2}}$ as well as the field renormalization $Z_{2}$
become \emph{negative} at some critical value of the coupling
constant. These critical values are $\lambda \approx 0.062$ corresponding
to $g/m_{2} \approx 3.13$ for $Z_{m_{2}}$ and $\lambda \approx 0.086$
corresponding to $g/m_{2} \approx 3.68$ for $Z_{2}$. 
Since the propagator for  particle 2 enters the BS equation we conclude that
the domain of validity of  the system of equations (\ref{eq: CJT BS expl. 1P}),
(\ref{eq: CJT DSE expl.12P}),  (\ref{eq: CJT DSE expl.3P}) is limited to the
range $0 \leq g/m_{2} \leq 3.13$. A negative value for $Z_{m_{2}}$ especially
implies that $m_{2}^2 < 0$, i.e.\ particle 2 becomes tachyonic.
Note also that in ref. \cite{RosSchr} the
critical value $\lambda_{c} \approx 0.063$ of the coupling constant was found
using a variational approach.

It remains to solve the BS equation thereby employing the dressed propagators
which are the solution of (\ref{eq: CJT DSren 1,2}) and (\ref{eq: CJT DSren
3}). Restricting the calculations to the allowed range of the coupling constant
$g$ one does find that the inclusion of the self-energies somewhat lowers the
solutions $g(M)$, however, the deviations are much smaller than $1\%$. Leaving
the domain of validity i.e.\ going beyond $g/m_{2} \approx 3.68$ one does find a
growing influence of the self-energies as could be expected. For $g/m_{2} \gg
3.68$ we also found complex eigenvalues $\lambda$.

The effect of the inclusion of the set of crossed ladders has been
investigated by Nieuwenhuis and Tjon \cite{NieuTjon}.
They used the Feynman-Schwinger representation which takes account
of the ordinary as well as the crossed ladders. Within this framework they
calculated the binding energy of the ground state and compared with the
corresponding results of the various bound-state equations. They concluded
that the ladder Bethe-Salpeter equation substantially underestimates the binding
energy of the ground state for large values of the coupling constant.
This seems to support the results of this work. However, we caution the
reader not to compare the results directly. Nieuwenhuis and Tjon calculated
the effect of the inclusion of the crossed ladders and used bare propagators
within the Bethe-Salpeter equation as well as within the Feynman-Schwinger
representation. This may be used to estimate the importance of
the diagrams that have been neglected in the kernel of the ladder Bethe-Salpeter
equation.
The present work emphasizes the necessity to use consistent
approximations for the kernel of the Bethe-Salpeter equation and for the
self-energies of all particles. It has been shown that the renormalizability
of the one-particle
properties restricts the domain of validity of the approximation that has
been used for the effective action.
Both approaches limit the applicability of the ladder approximation,
but they do so using quite different constraining principles.

To summarise: the physical condition $0 \leq Z_i,Z_{m_{i}} \leq 1$ defines a domain of
validity for the coupled system of Dyson-Schwinger equations (\ref{eq: CJT
DSren 1,2}) and (\ref{eq: CJT DSren 3}) and the solution of this system of
equations gives the propagators that are to be used in the BS equation
(\ref{eq: CJT BS expl. 1P}). Therefore the condition $0 \leq Z_i,Z_{m_{i}} \leq 1$ gives
an upper bound for the coupling constant $g$ above which the quanta of the
field $\Phi_{2}$ no longer correspond to physical particles and above which the
ladder-approximation to the BS equation is not applicable.

\section{Bound States in QED}
\label{sec. BSE QED}

It is certainly interesting to compare the results obtained so far in the
scalar model with an exactly renormalizable theory. QED whose Dyson--Schwinger
and BS equations have been studied intensiveley \cite{Rob94,Miranski} provides
hereby a good testing ground for interpreting our results. We first calculate
the spectrum of the ladder approximation to the BS equation for positronium,
and we will compare to the results that have been obtained for the scalar
theory. As in the last section we will derive consistent Dyson-Schwinger and BS
equations, taking one approximation for the CJT-action as starting point. A
simplified version of the coupled system of Dyson Schwinger equations will be
solved.

First, we will discuss the BS equation in ladder approximation
for positronium. However, we will also present results for constituents of
different mass and for a massive photon when commenting on the problem
of complex eigenvalues. The ladder BS equation is given by
\begin{eqnarray}
\Gamma \left( q,P\right) &=&
-ie^{2}\int
\frac{d^{4}k}{\left( 2\pi \right) ^{4}}\,D^{\mu \nu }\left( q-k\right)
\gamma _{\mu }S_{1}\left( k_{+} \right)
\Gamma \left( k,P\right) S_{2}\left( k_{-} \right) \gamma _{\nu } \\
&& \quad\quad\quad\quad\quad\quad\quad k_{+} = k + \alpha P  \nonumber\\
&& \quad\quad\quad\quad\quad\quad\quad k_{-} = k - (1-\alpha) P \nonumber
\label{eq: QED BSE allg}
\end{eqnarray}
where $\Gamma$ is the BS vertex function which is related to the BS amplitude
as given in (\ref{X1.37}). The propagators of the constituents are approximated
by the corresponding bare propagators
\begin{equation}
   S_{i}\left( p\right) = \frac{p\hspace{-0.2cm}/ +m_{i}}{
   p^{2}-m_{i}^{2}+i\epsilon },\quad i=1,2.
\end{equation}
$D^{\mu \nu }$
is the bare propagator of the photon, and since we are working in the
ladder approximation we used a bare fermion--photon  vertex, i.e.\
$\gamma_{\mu}$ only.

The general expression for the BS vertex function $\Gamma$ describing a
pseudoscalar bound state was derived in \cite{Lewll69}. However, in
the Feynman gauge and considering only the ladder approximation one
can derive that the tensor part of the vertex function vanishes.
Taking this into account we arrive at
\begin{equation}
\Gamma \left( q,P\right) =\gamma _{5}\biggl( \Gamma _{1}\left(
q,P\right) +\left( P\cdot q\right) q\hspace{-0.2cm}/\, \Gamma
_{2}\left( q,P\right) +P\!\hspace{-0.20cm}/ \,\,\Gamma _{3}\left(
q,P\right) \biggr)  \label{X6.8}
\end{equation}
Substituting this ansatz into the BS equation (\ref{eq: QED BSE allg})
and taking appropriate combination of traces and projections one
arrives at a coupled system of integral equations for the scalar functions
$\Gamma_{1},\Gamma_{2}$ and $\Gamma_{3}$. The details of the numerical
method are given in appendix \ref{app. BSE QED}, and we proceed immediately to the
discussion of the results.

\begin{figure}
\epsfig{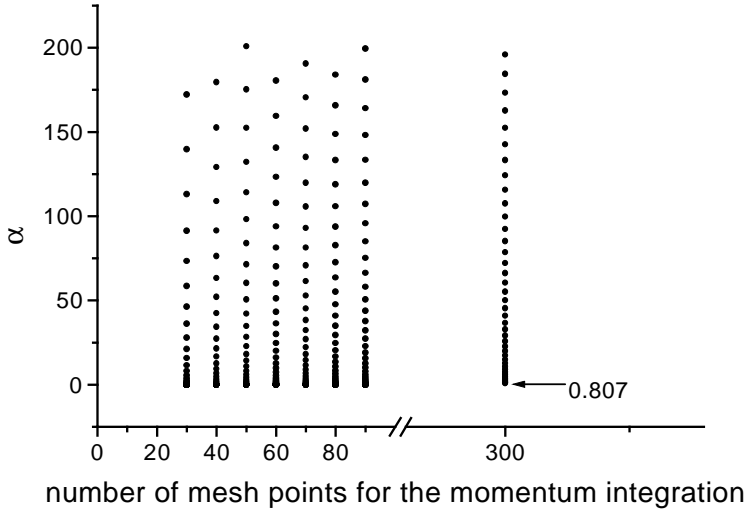}
\caption{ Solutions of the BS equation (\ref{eq: QED BSE allg}) for
          constituents of equal mass in the limiting case
          $M \rightarrow 0$. The solutions are given for
          $0 \leq \alpha \leq 200$ and for various numbers of meshpoints.}
\label{Abb71}
\end{figure}

First, we consider the spectrum for constituents of equal mass in the limit
of massless bound states. Goldstein \cite{Goldstein} was the first who
considered this limiting case. He found massless solutions for
\emph{all} values of the coupling constant. One can demonstrate that the
BS equation (\ref{eq: QED BSE allg}) in this limit reduces to an
equation for $\Gamma_{1}$ which is not coupled to
$\Gamma_{2}$ and $\Gamma_{3}$. Therefore one may impose the
``normalizibility constraint''
\begin{equation}
  \int d^{4}k \left| \Gamma_{1}(k) \right| < \infty .
\end{equation}
Taking this condition into account Goldstein still found a continuum
of solutions which, however, starts at $\alpha = e^{2}/4\pi=\pi/4$.
The results of the numerical solution are displayed in fig.\ \ref{Abb71}
for an increasing number of mesh points for the numerical momentum integration.
Hereby every point represents a solution of the BS equation
for $M=0$ (within numerical accuracy), and the corresponding value of the fine
structure constant $\alpha$ is represented. As the number of
mesh points is increased one finds an ever increasing density of solutions
in the intervall $0 \leq \alpha \leq 200$. This is exactly the way
a continuum of solutions is expected to show up in a numerical treatment.
However, the lower bound of this continuum depends on the number of
mesh points for the momentum integration. Taking 40 meshpoints the
continuum starts at $\alpha \approx 0.59$ and taking as much as
300 meshpoints we found the lower bound to stabilize at
$\alpha \approx 0.807$. This compares reasonably well to $\alpha=\pi/4 \approx
0.785$ which was found by Goldstein. Note that this continuum of solutions is
not related to the physical continuum of the scattering states.

\begin{figure}[t]
\epsfig{file=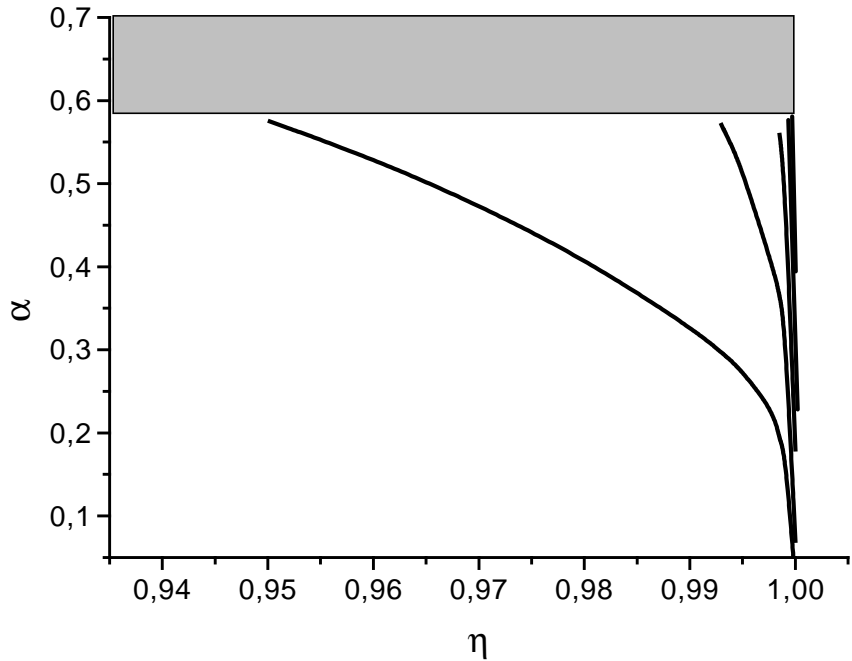,width=10cm,angle=-90}
\caption{ Solutions of the BS equation (\ref{eq: QED BSE allg}) for
          constituents of equal mass. The parameters are defined as
          $\alpha = e^{2}/4\pi$ and $\eta = M/(m_{1} +m_{2})$. The
          lines represent discrete solutions and the shaded area
          stands for a continuum of solutions. We took 40 meshpoints
          for the momentum-integration which lead to a lower bound
          of $\alpha \approx 0.59$ for the continuum.}
\label{Abb74}
\end{figure}

We turn to the case of rather weakly bound states of the electron
and the positron (weakly bound equal mass case). The results are
shown in fig. \ref{Abb74}. The lines represent discrete solutions which
start at $\alpha \approx 0$ as well as at greater values of $\alpha$
and the shaded area marks the beginning of a continuum of solutions. Since
we took 40 meshpoints for the standard calculations the continuum starts
at $\alpha \approx 0.59$.

For $0 < \eta \leq 0.94$ the continuum is the \emph{only} feature of the
spectrum
of solutions of the BS equation (\ref{eq: QED BSE allg}), i.e.\ the continuum of
solutions (which was known to appear for $M \rightarrow 0$) is present for
\emph{all} values of $\eta$, and it is so with a constant lower bound.

The BS equation for scalar constituents and scalar exchange particle gave
complex solutions $\alpha(\eta)$ for constituents of different mass and for a
massive exchange particle. We solved the BS equation (\ref{eq: QED BSE allg})
for various mass ratios $m_{1}/m_{2} \neq 1$ and $m_{\gamma}/m_{2} \neq 0$ and
the results agree qualitatively with those that have been obtained for the
scalar theory.
To mention two of the more pronounced features: a massive photon destroys
the clustering of solutions at
$\alpha = 0$ and for $M \rightarrow m_{1}+m_{2}$. As well as for the scalar
theory we found complex solutions $\alpha(\eta)$ which, however, were found
only \emph{above} the lower bound of the continuum.

\begin{figure}
\begin{center}
$\biggl(
\begin{minipage}{20mm}
  \epsfig{file=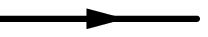,width=20mm}
\end{minipage}
\biggr)^{-1}
=\hspace{4mm}
\biggl(
\begin{minipage}{20mm}
  \epsfig{file=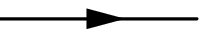,width=20mm}
\end{minipage}
\biggr)^{-1}
+\hspace{4mm}
\begin{minipage}{2cm}
  \epsfig{file=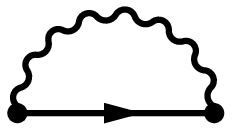,width=2cm}
\end{minipage}
$
\end{center}
\vspace{-2cm}
\begin{center}
$\biggl(
\begin{minipage}{20mm}
  \epsfig{file=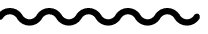,width=20mm}
\end{minipage}
\biggr)^{-1}
=\hspace{4mm}
\biggl(
\begin{minipage}{20mm}
  \epsfig{file=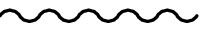,width=20mm}
\end{minipage}
\biggr)^{-1}
+\hspace{4mm}
\begin{minipage}{2.0cm}
  \hspace{0.2cm}
  \epsfig{file=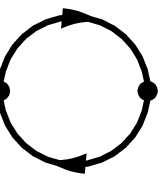,width=1.4cm}
\end{minipage}
$
\end{center}
\caption{ Graphical representation of the coupled system of Dyson-Schwinger
          equations
         (\ref{X7.1}). Thick and thin lines represent the dressed and the
         bare propagators,
         respectively. }
\label{DSE_123_QED}
\end{figure}

Let us finally take up the discussion of the (in-)consistency of the
approximations
for the propagators and for the kernel that enter the BS equation
(\ref{eq: QED BSE allg}). As for the scalar theory we approximate the
loop series $\Gamma_{2}$ by
\begin{eqnarray}
\Gamma_{2} &=&
\begin{minipage}{12mm}
 \epsfig{file=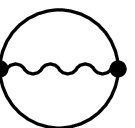,width=10mm}
\end{minipage}
\nonumber \\
&=& \frac{e^{2}}{2}\int d^{4}x\,d^{4}y\,{\mathrm{tr}} \biggl( S\left(
x,y\right) \gamma ^{\mu }S\left( y,x\right) \gamma ^{\nu }D_{\mu \nu }\left(
x,y\right) \biggr)
\label{eq: QED G2 App}
\end{eqnarray}
where $S$ and $D_{\mu \nu }$ are the propagators of the electron and the photon,
respectively.
These propagators are not the corresponding bare propagators but the solution of
the coupled system of Dyson-Schwinger equations:
\begin{equation}
\frac{\delta \Gamma }{\delta S}=\frac{\delta \Gamma }{\delta D_{\mu \nu }}=0 .
\label{X7.1}
\end{equation}
Using the approximation (\ref{eq: QED G2 App}) for $\Gamma_{2}$ one may derive
the explicit Dyson--Schwinger equations that correspond to (\ref{X7.1}). This
leads to a system of equations which is shown in fig. \ref{DSE_123_QED} and
which compares to the system of equations for the scalar theory fig.
\ref{DSE_123_boson}.

If the appoximation (\ref{eq: QED G2 App}) is used as input for the general form
of the BS equation
\begin{equation}
\int d^{4}x^{\prime }\,d^{4}y^{\prime }\,\Gamma _{ik,i^{\prime }k^{\prime
}}^{\left( 2\right) }\left( x,y;x^{\prime },y^{\prime }\right) \psi
^{i^{\prime }k^{\prime }}\left( x^{\prime },y^{\prime }\right) =0
\label{eq: CJT allg. BSE 2 QED}
\end{equation}
one obtains the ladder approximation to the BS equation which is given in
(\ref{eq: QED BSE allg}). However, it has to be emphasized that the propagators
$S$ and $D$ are not the bare ones but the solution of (\ref{X7.1}).

For the scalar theory we showed that the physical condition $0 \leq Z_i,Z_{m_{i}} \leq 1$
gives a domain of validity for the system of Dyson--Schwinger equations as well
as for the ladder approximation to the BS equation. In order to obtain a first
estimate we solved the system of equations in quenched approximation thus
effectively neglecting the equation for the photon. The equation for the
propagator of the electron is then given by
\begin{equation}
S^{-1}\left(
p\right) =Z_{2}\left( p\hspace{-0.2cm}/ -m_{0}\right)
-iZ_{1}e^{2}\int^{\Lambda }\frac{d^{4}k}{\left( 2\pi \right) ^{4}}\gamma ^{\mu
}S\left( k\right) \gamma ^{\nu }D_{\mu \nu }\left( k-p\right)
\label{eq: QA DSE}
\end{equation}
where the bare mass $m_{0}$ as well as the renormalization constants $Z_{1}$
and $Z_{2}$ depend on the renormalization scale $\mu$ and on the cutoff
$\Lambda$. As one can read off from (\ref{eq: QA DSE}) the approximation
(\ref{eq: QED G2 App}) yields the bare fermion--photon vertex together with an
obviously nonperturbative propagator for the fermion. This implies that the
Ward--Takahashi identity cannot be satisfied, i.e.\ $Z_{1} = Z_{2}$ cannot be
assumed. Since the use of the bare vertex enforces $Z_{1} = 1$ we determined
only $Z_{2}$ and $m/m_{0}$ according to the renormalization condition
\begin{equation}
\left. S^{-1}\left( p\right) \right|
_{p^{2}=0}=p\hspace{-0.2cm}/ -m
\label{X7.6}
\end{equation}
and fixed $Z_{1}$ to 1. For simplicity we employed the normalization at the
soft point $p^{2}=0$ instead of the on--shell renormalization.
Defining the scalar functions $A$ and $B$ through the relation
\begin{equation}
S\left( p\right) =\frac{1}{p\hspace{-0.2cm}/ A\left(
p^{2}\right) -B\left( p^{2}\right) }
\label{eq: QED Def AuB}
\end{equation}
one obtains from (\ref{eq: QA DSE}) a coupled system of equations for $A$ and
$B$ after taking appropriate combination of traces. In terms of these scalar
functions the renormalization conditions read $A(0)=1$ and $B(0)=m$.

The details of the numerical solution are given in appendix \ref{app. DSE QED}, 
and therefore we
focus on the results. For the physical coupling constant $\alpha = 1/137$ the
scalar functions $A$ and $B$ are nearly equal to their perturbative limits, i.e.\
$A \approx 1$ and $B \approx m$, as could be anticipated.

\begin{figure}
\epsfig{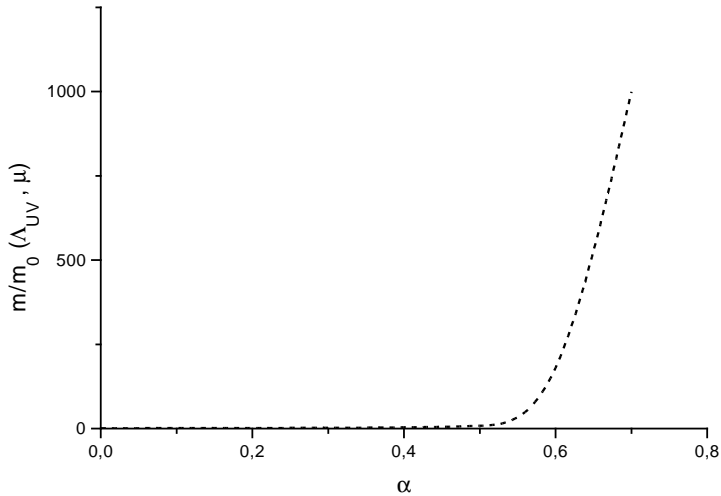}
\caption{ The mass--ratio $m/m_{0}$ versus the coupling constant $\alpha$.
          The divergence of this quantity signals chiral symmetry breaking.}
\label{mm0}
\end{figure}

\begin{figure}
\epsfig{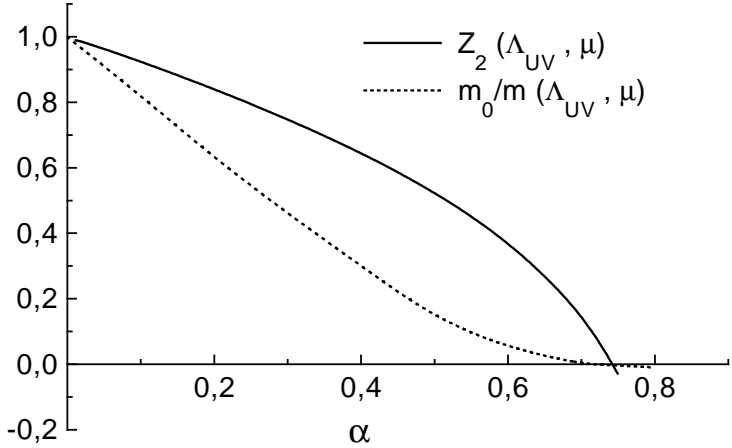}
\caption{ The renormalization constant $Z_2$ and the mass-ratio $m_{0}/m$
          versus the coupling constant $\alpha$.}
\label{Abb83}
\end{figure}

For greater values of $\alpha$ an increasing number of iterations was needed in
order to obtain a self--consistent solution. We found no solution for
$\alpha > 0.74$ which compares quite well with the critical value
$\alpha = \pi/4 \approx 0.785$ and with the lower bound of the continuum, which
was estimated to be at $\alpha \approx 0.807$. Fig.\ \ref{mm0} displays the
mass--ratio $m/m_{0}$. The divergence of this quantity signals dynamical
generation of a mass for vanishing bare mass, i.e.\ spontaneous breaking of
chiral symmetry. To clarify the significance of this result in comparison to
the scalar model we plot in fig.\ \ref{Abb83} the inverse mass--ratio $m_{0}/m$
and the renormalization constant $Z_{2}$ versus the
coupling constant $\alpha$. Both $Z_{2}$ and $m_{0}/m$ change sign at the same
value of $\alpha$,
$\alpha \approx 0.75$. A commonly accepted interpretation of this is that
dynamical chiral symmetry breaking occurs at the critical value of the coupling
constant, and that above this value of the renormalized fine structure
constant QED is not defined \cite{Miranski,Haw97}.

\section{Conclusions}

The central aim of this investigation was a clarification how (or more
precisely, whether) the physical spectrum of excited states may be infered
from the fully relativistic Bethe--Salpeter equation. In a very strict sense
the answer is negative: Taking into account the need for renormalization in the
underlying theory the applicability of the ladder Bethe--Salpeter equation
turns out to be very restricted. The comparison to (quenched) QED demonstrates
this clearly. It is asserted that the value of renormalized fine structure
constant
has to be lower than some critical value in order to allow for reasonably
behaved quantum field theory, see e.g.\ \cite{Haw97} for a corresponding result
in a Dyson-Schwinger
approach and \cite{Goe} for a lattice calculation.
Finding the similar behaviour that the field renormalization constants becomes
negative for large couplings one may speculate that our (super--renormalizable)
scalar model is also well--defined only for renormalized couplings below the
critical one signaled by the zero of the renormalization constant.
Below these critical values of the coupling constant the
Dyson--Schwinger equations for the
propagators as well as the Bethe--Salpeter equation in rainbow--ladder
approximation were shown to give physically acceptable solutions
only. In addition, the use of bare propagators in the Bethe--Salpeter equation
introduces only a small additional error.

For large couplings the homogeneous ladder Bethe--Salpeter equation possesses
abnormal solutions being excitations in the relative time coordinate. Half of
these abnormal states are odd w.r.t. this relative time,
and they lead to complex eigenvalues via the crossing with even states. In
addition, in QED there is the problem of a continuum of solutions
(Goldstein problem). The investigation  reported here provides evidence
that all these crazy things happen beyond the applicability region of the
underlying renormalized quantum field theory. Thus, before studying a
spectrum using the Bethe--Salpeter equation one should ensure oneself that
the renormalization of one--particle properties can be performed with a
physically acceptable result.

\bigskip
\bigskip
\centerline{\Large\bf Acknowledgments}
\smallskip
We thank Gerhard Hellstern for discussions and contributions in the
early stages of this work and Martin Oettel for a critical reading
of the manuscript and his comments.
Furthermore, we thank Lorenz von Smekal for helpful discussions.
We are grateful to Hugo Reinhardt for encouragement and support.

This work has been supported by BMBF under contract 06TU888.

\newpage
\appendix{}
\section*{Appendix}

\section{Numerical method for the solution of the scalar Bethe--Salpeter
equation}
\label{App. BSE1}

The Wick--rotated homogeneous BS equation for scalar constituents and
scalar exchange particle is given by
\begin{equation}
\chi \left( \bar{p},\bar{P}\right) =g^{2}\frac{1}{\bar{p}_{1}^{2}+m_{1}^{2}}%
\frac{1}{\bar{p}_{2}^{2}+m_{2}^{2}}\int \frac{d^{4}\bar{p}^{\prime }}{\left(
2\pi \right) ^{4}}\frac{\chi \left( \bar{p}^{\prime },\bar{P}\right) }{%
\left( \bar{p}-\bar{p}^{\prime }\right) ^{2}+\mu ^{2}}\, ,  \label{X9.3}
\end{equation}
where the euclidean momenta $\bar{p}_{i}$ are defined according to
\begin{equation}
  \bar{p}_{1} := \alpha \bar{P} + \bar{p} , \quad
  \bar{p}_{2} := (1-\alpha) \bar{P} - \bar{p}.
\end{equation}
The symmetries of this equation have been discussed in section \ref{sec. sym.} and
are summarised in table \ref{t1}. For a massless exchange particle
(\ref{X9.3}) is invariant under $O(4)$ transformations and the BS
amplitudes are proportional to a spherical harmonic of $SO(4)$,
denoted by $Z_{klm}$. In the case of $O(3)$ angular momentum $l=0$
the $Z_{klm}$ are proportional to Gegenbauer polynomials $C_{k}^{1}$ of degree 1,
see \cite{Abram} for their definition.
For a massive exchange particle the $O(4)$-symmetry is only approximate.
Nevertheless, it turns out that the expansion
\begin{equation}
 \chi \left( \bar{p},\bar{P}\right) =\sum_{m=0}^{\infty }C_{m}^{1}\left( \cos
 \left( \theta \right) \right) \chi _{m}\left( \left\| \bar{p}\right\|
 ,\left\| \bar{P}\right\| \right) \quad \quad \theta
 :=\sphericalangle \left( \bar{p},\bar{P} \right)
 \label{X9.5}
\end{equation}
still converges very fast.
For further usage we now define
the dimensionless ``momenta''
\begin{equation}\label{def. dimless 1}
  x       :=   \frac{\| \bar{p} \|}{m_{2}} ,             \quad
  x'      :=   \frac{\| \bar{p}' \|}{m_{2}} ,            \quad
\end{equation}
the dimensionless mass-parameters
\begin{equation}\label{def. dimless 2}
  \delta  :=   \frac{m_{1}}{m_{2}}   ,                   \quad
  \kappa  :=   \frac{\mu}{m_{2}}     ,                   \quad
  \eta    :=   \frac{M}{m_{1}+m_{2}} ,
\end{equation}
and the dimensionless coupling constant
\begin{equation}\label{def. dimless 3}
  \lambda :=   \frac{g^2}{16\pi^2 m_{2}^{2}} \, .
\end{equation}
Furthermore the abbreviations
\begin{equation}
  z      :=   \frac{\kappa^{2} + x^{2} + x'^{2}}{2xx'} ,\quad
  B      :=   z-\sqrt{z^{2}-1}
\end{equation}
will be used.
The integration over the euclidean momentum $\bar{p}'$
is done in spherical coordinates:
\begin{equation}
  \int d^{4}\bar{p}^{\prime }=\int_{0}^{\infty }d\bar{p}^{\prime }\bar{p}
  ^{\prime 3}\int_{0}^{\pi }d\theta ^{\prime }\sin ^{2}\theta ^{\prime
  }\int_{0}^{\pi }d\varphi ^{\prime }\sin \varphi ^{\prime }\int_{0}^{2\pi
  }d\psi ^{\prime } \, ,
  \quad\quad \theta' := \sphericalangle \left( \bar{p}',\bar{P} \right).
\end{equation}
The integrations over $\theta' , \varphi' , \psi'$ can be done
analytically and the integral over the absolute value of $\bar{p}'$
has been mapped to the interval $[-1,+1]$ according to
\begin{equation}\label{int.trans.}
  x' = \tan(\frac{\pi}{4}(1+z))
  \quad
  \int_{0}^{\infty }dx^{\prime }\rightarrow \frac{\pi }{4}\int_{-1}^{+1}\frac{
  dz}{\cos ^{2}\left( \frac{\pi }{4}\left( 1+z\right) \right) }  \label{X9.22}
\end{equation}
for the numerical $z$-integration
\begin{equation}
  \int_{-1}^{+1} dz f(z) 
  \quad \rightarrow \quad
  \sum_{j=1}^{N_{1}} w_{z}(j) f(z_{j}).
\end{equation}
$w_{z}(j)$ are hereby the weights of the Gauss-Legendre integration.
After projection onto the expansion coefficients $\chi_{k}$ we arrive
at an eigenvalue problem for the coupling constant
$\lambda$
\begin{equation}\label{BSE SS ewsys}
  \phi _{k}\left( x_{i},\eta \right) =\lambda
  \sum_{n=0}^{N_{2}}\sum_{j=1}^{N_{1}}A_{ki,nj}\cdot \phi _{n}\left(
  x_{j},\eta \right) \, ,
\end{equation}
where the auxiliary functions $\phi_{k}:=i^{k} x^{2} \chi_{k} (\bar{p} , \bar{P})$
have been used. The indices $n$ and $i$ run over the degree of the Gegenbauer
polynomial and over the $N_{1}$ mesh points for the momentum integration,
respectively. The definition of the matrix
$A_{ki,nj}$ is given by
\begin{eqnarray}
  A_{ki,nj}  &:=& \frac{\pi }{4}\frac{1}{\delta \eta ^{2}x_{i}}\frac{1}{y_{1}+y_{2}}
                \frac{w_{z}\left( j\right) }{\cos ^{2}\left( \frac{\pi }{4}\left(
                1+z_{j}\right) \right) }\frac{\left[ -i^{k-n-1}S_{nk}\right] }
                {n+1}B^{n+1}, \label{matrix def.} \\
  y_{1}      &:=& \frac{\delta \left( 1-\eta ^{2}\right) +x_{i}^{2}}{2\delta \eta x_{i}},
  \quad \quad
  y_{2}      := \frac{1-\eta ^{2}+x_{i}^{2}}{2\eta x_{i}}, \\
  S_{nk}     &:=& \frac{\left( -1\right) ^{n}}{i^{k-n-1}}\left( \frac{\left(
  -1\right) ^{\sigma +k+n}A_{1}^{\left| n-k\right| +1}+A_{1}^{n+k+3}}{1+A_{1}^{2}
  }\right. , \\
  &\quad& \quad \quad \quad \quad \quad +\left. 
  \frac{\left( -1\right) ^{\sigma }A_{2}^{\left| n-k\right| +1}-\left( -1\right)
  ^{n+k+1}A_{2}^{n+k+3}}{1+A_{2}^{2}}\right), \\
  \sigma  &:=& \frac{1}{2}\left( \left| n-k\right| +k+n\right), \\
  &{\rm and}& \\
  A_{i}      &:=& y_{i}-\sqrt{y_{i}^{2}+1} \, .
\end{eqnarray}
Different LAPACK-routines have been used to solve the
eigenvalue problem (\ref{BSE SS ewsys}) and to cross-check the results.

\section{The Bethe--Salpeter amplitudes for euclidean momenta an for constituents
of equal mass}
\label{app. relat.}

The BS amplitudes describing a bound state of scalar constituents of equal mass
are defined by
\begin{eqnarray}
  \chi \left( x_{1},x_{2},P\right) &=&\left\langle 0\left| T\left\{ \Phi
  \left( x_{1}\right) \Phi \left( x_{2}\right) \right\} \right| P\right\rangle \\
  \overline{\chi }\left( x_{1},x_{2},P\right) &=& \left\langle 0\left| T\left\{ \Phi^{\dagger}
  \left( x_{1}\right) \Phi^{\dagger} \left( x_{2}\right) \right\} \right| P\right\rangle \\
  &=&\left\langle 0\left| 
  \overline{T}\left\{ \Phi \left( x_{1}\right) \Phi \left( x_{2}\right)
  \right\} \right| P\right\rangle ^{*} \nonumber
\end{eqnarray}
where $T$ is the usual time ordering operator and
$\overline{T}$ denotes time ordering in the reverse order.
Now we use the definitions
\begin{eqnarray}
  X &=& \alpha x_{1} + (1-\alpha) x_{2} \\
  x &=& x_{1} - x{2} \nonumber
\end{eqnarray}
to separate the center of mass motion
\begin{equation}
  \chi \left( x_{1},x_{2};P\right) =e^{-iP\cdot X}\,\phi \left( x,P\right)
\end{equation}
defining thereby $\phi$, which denotes the relative part of the BS amplitude.
Taking this separation of the center of mass and writing the time
ordering explicitely we obtain
\begin{eqnarray}
\chi \left( x_{1},x_{2},P\right) &=& e^{-iP\cdot X}\left( \Theta \left(
x^{0}\right) g\left( x,P\right) \,+\Theta \left( -x^{0}\right) g\left(
-x,P\right) \,\right)  \label{X4.12} \\
\overline{\chi }\left( x_{1},x_{2},P\right) &=& e^{iP\cdot X}\left( \Theta
\left( x^{0}\right) g\left( -x,P\right) \,+\Theta \left( -x^{0}\right)
g\left( x,P\right) \right) ^{*}  \, .\label{X4.13}
\end{eqnarray}
Since we assumed constituents of equal mass we may take the amplitudes
to have a definite $x_{0}$-parity $\epsilon \left( \phi \right)$
which is defined by
\begin{equation}
\phi \left( -x^{0},\mathbf{x}\right) =\epsilon \left( \phi \right) \,\phi
\left( x^{0},\mathbf{x}\right) \, .  \label{X4.14}
\end{equation}
This can be used to rewrite eqs. (\ref{X4.12}),(\ref{X4.13}) in the
following way
\begin{eqnarray}
\phi \left( x_{1},x_{2},P\right) &=&\left( \Theta \left(
x^{0}\right) g\left( x,P\right) \,+\Theta \left( -x^{0}\right) g\left(
-x,P\right) \,\right) \, , \label{X4.16} \\
\overline{\phi }\left( x_{1},x_{2},P\right) &=&\,\epsilon
\left( \phi \right) \,\left( \Theta \left( x^{0}\right) g\left( -x^{0},%
\mathbf{x},P\right) \,+\Theta \left( -x^{0}\right) g\left( x^{0},-\mathbf{x}%
,P\right) \right) ^{*} \, . 
\end{eqnarray}
Using the relation
\begin{equation}
\Theta \left( z\right) =-\frac{1}{2\pi i}\int dk\,e^{-ikz}\,\frac{1}{k+i0^{+}}
\end{equation}
we find the following Fourier-transforms
\begin{eqnarray}
\int d^{4}x\,e^{-ip\cdot x}\Theta \left( x^{0}\right) g\left( x,P\right) &=&-%
\frac{1}{2\pi i}\int dq^{0}\,\frac{\widetilde{g}\left( q^{0},\mathbf{p}%
,P\right) }{q^{0}-p^{0}+i0^{+}} \, , \label{X4.19} \\
\int d^{4}x\,e^{-ip\cdot x}\Theta \left( -x^{0}\right) g\left( -x,P\right)
&=&-\frac{1}{2\pi i}\int dq^{0}\frac{\widetilde{g}\left( q^{0},-\mathbf{p}%
,P\right) }{q^{0}+p^{0}+i0^{+}} \, , \nonumber \\
\int d^{4}x\,e^{-ip\cdot x}\Theta \left( x^{0}\right) \left[ g\left( -x^{0},%
\mathbf{x},P\right) \right] ^{*} &=&-\frac{1}{2\pi i}\int dq^{0}\frac{%
\widetilde{g}^{*}\left( q^{0},-\mathbf{p},P\right) }{q^{0}-p^{0}+i0^{+}} 
\nonumber\, , \\
\int d^{4}x\,e^{-ip\cdot x}\Theta \left( -x^{0}\right) \left[ g\left( x^{0},-%
\mathbf{x},P\right) \right] ^{*} &=&-\frac{1}{2\pi i}\int dq^{0}\frac{%
\widetilde{g}^{*}\left( q^{0},\mathbf{p},P\right) }{q^{0}+p^{0}+i0^{+}} 
\nonumber
\end{eqnarray}
which may be used to rebuild the Fourier transforms of the relative part
$\phi$ of the BS amplitudes.
After Wick-rotation $p_{0} \rightarrow ip_{4}$ we find the relation
\begin{equation}
 \widetilde{\overline{\phi }}\left( \bar{p},P\right) =\epsilon \left( \phi
\right) \left[ \widetilde{\phi }\left( \bar{p},P\right) \right] ^{*}
\end{equation}
for the relative parts of the BS amplitudes and for euclidean relative
momenta $\bar{p}$.

\section{Numerical method for the solution of the Dyson-Schwinger equations of
         the scalar theory}
\label{app. SS DSE}

The coupled system of DS equations for the propagators of the scalar fields
$\Phi_{i}\, , i=1,2,3$, is given by
\begin{eqnarray}
\Sigma _{i}\left( p^{2}\right) &=& \Delta_{i} +Z_{i}
2g^{2}i\int \frac{d^{4}q}{\left( 2\pi \right) ^{4}}\,G_{3}\left( p-q\right)
G_{i}\left( q\right)\, , \quad \quad i=1,2 \, ,
\label{eq: CJT DSren 1,2 app} \\
\Sigma _{3}\left( p^{2}\right) &=& \Delta_{3}  +%
g^{2}i\int \frac{d^{4}q}{\left( 2\pi \right) ^{4}}\,
\sum_{i=1}^{2} Z_{i}
G_{i}\left( p-q\right) G_{i}\left( q\right)
\label{eq: CJT DSren 3 app}
\end{eqnarray}
where the abbreviation
\begin{equation}
  \Delta _{i}:=\left( 1-Z_{i}\right) p^{2}+\left(
  Z_{i}Z_{m_{i}}-1\right) m_{i}^{2} \, , \qquad i=1,2,3 \, ,
\end{equation}
and the definition
$\Sigma_{i}(p^{2}):=p^{2}-m_{i}^{2}-G_{i}^{-1}(p)$
have been used. For notational convenience we introduce
\begin{equation}
  F_{ik}(p^{2})  :=  \int \frac{d^{4}q}{(2\pi)^4} G_{i}(p-q) G_{k}(q) \, ,
                       \quad\quad i,k=1,2,3\, ,
\end{equation}
which allows to rewrite the equations (\ref{eq: CJT DSren 1,2 app}),(\ref{eq: CJT DSren 3 app})
in the form
\begin{eqnarray}
  \Sigma _{i}\left( p^{2}\right) &=& \left( 1-Z_{i}\right) p^{2}+\left(
                                     Z_{i}Z_{m_{i}}-1\right) m_{i}^{2}
                                     + Z_{i}2g^{2}i F_{3i}(p^{2}) \, ,
                                     \quad\quad i=1,2 \, , \\
  \Sigma _{3}\left( p^{2}\right) &=& \left( 1-Z_{3}\right) p^{2}+\left(
                                     Z_{3}Z_{m_{3}}-1\right) m_{3}^{2}
                                     + g^{2}i \sum_{i=1}^{2} Z_{i} F_{ii}(p^{2}) \, .
\end{eqnarray}
We use the on-shell renormalization condition
$\Sigma _{i}\left( m_{i}^{2}\right) = 0$ to eliminate $Z_{m_{i}}$
and obtain
\begin{eqnarray}
  \Sigma _{i}\left( p^{2}\right) &=& (1-Z_{i})(p^{2}-m_{i}^{2}) + Z_{i}2g^{2}i
                                     \biggl( F_{3i}(p^{2}) - F_{3i}(m_{i}^{2})\biggr) \, ,
                                     \quad i=1,2 \, , \\
  \Sigma _{3}\left( p^{2}\right) &=& (1-Z_{3})(p^{2}-m_{3}^{2}) + g^{2}i
                                     \sum_{i=1}^{2} Z_{i}
                                     \biggl( F_{ii}(p^{2}) - F_{ii}(m_{3}^{2})\biggr) \, .
\end{eqnarray}
The $F_{ik}$ contain the self--energies $\Sigma_{i} \left( (p-q)^2 \right)$ and
$\Sigma_{k} \left( q^2 \right)$. In the following we use an angle approximation,
i.e.\ we substitute $\max(q^2,p^2)$ for $(p-q)^2$ if (and only if) $(p-q)^2$ appears as
the argument of $\Sigma_{i}\, , \, i=1,2,3$. After Wick-rotation
\begin{equation}
q = (q_{0},\mathbf{q}) \rightarrow (iq_{4},\mathbf{q})=: \bar{q}
\end{equation}
we change to spherical coordinates
\begin{equation}
  \int d^{4}\bar{q}       \rightarrow
  \int_{0}^{\Lambda_{UV}} d\bar{q}\bar{q}^{3}
  \int_{0}^{\pi}          d\theta\sin ^{2}\theta
  \int_{0}^{\pi}          d\varphi\sin \varphi
  \int_{0}^{2\pi}         d\psi
\end{equation}
where we temporarily introduced the cutoff $\Lambda_{UV}$. Because of the angle approximation
the integrations over $\theta,\varphi$ and $\psi$ can be done analytically
which leads to
\begin{equation}
  F_{ik}      = \frac{ i }{ 32\pi^{2} p^{2} }
                \int_{0}^{\Lambda} d\bar{q}^{2}
                \frac{ \sqrt{c^{2} + 4\bar{p}^{2}\bar{q}^{2}} - c          }
                     { \bar{q}^{2} + m_{k}^{2} + \Sigma_{k}(-\bar{q}^{2})  } \, ,
\end{equation}
where the abbreviation
\begin{equation}
  c  :=  -\bar{p}^{2} + \bar{q}^{2} + m_{i}^{2}
         + \Sigma_{i}\left( -\max(\bar{p}^{2},\bar{q}^{2}) \right)
\end{equation}
has been used for convenience.
The system of equations has been solved iteratively and in each step of the iteration
the field renormalization constants $Z_{i}$ are calculated according to
\begin{equation}
  \left. \frac{d}{dp^{2}} \Sigma(p^{2}) \right| _{p^{2} = m^{2}} = 0 \, .
\end{equation}

The solutions $\Sigma_{i}\, ,\, i=1,2,3\, ,$ of the system of equations
(\ref{eq: CJT DSren 1,2 app}) and (\ref{eq: CJT DSren 3 app})
enter the BS equation via the propagators of the constituents and the
exchange particle, see equation (\ref{eq: CJT BS expl. 1P}) and
its derivation. More explicitely:
\begin{equation}
  \psi^{12} \left( p,P\right) = \alpha g^{2} i G_{1}\left( p
  _{1}\right) G_{2}\left( p_{2}\right) \int \frac{d^{4}q}{
  \left( 2\pi \right) ^{4}}\frac{\psi^{12} \left( q,P\right) }{\left( 
  p-q\right) ^{2}+m_{3}^{2}+\Sigma _{3}\left( \left( p-q
  \right) ^{2}\right) }  \label{eq: CJT Lösm. BSE app.}
\end{equation}
where
\begin{equation}
  p_{1} := p - \alpha P\, , \quad p_{2} := p + (1-\alpha) P
\end{equation}
and
\begin{equation}
G_{i}\left( p_{i}\right) =\frac{1}{p_{i}^{2}+m_{i}^{2}+%
\Sigma _{i}\left( p_{i}^{2}\right) }\, , \quad\quad i=1,2 \, .\label{X5.32}
\end{equation}
This equation is no longer an eigenvalue problem for the square of the
coupling constant $g^{2}$. However, one may introduce a formal eigenvalue
$\alpha$ and regard $g$ and $M$ as parameters. A solution, i.e.\ a pair
$(g,M)$ is now signaled by the eigenvalue $\alpha=1$.
Since we use the angle approximation it is sufficient to apply the
substitution
$m_{i}^{2} \rightarrow m_{i}^{2} + \Sigma_{i}(p^2_{\max})$
to the eigenvalue problem as given in (\ref{BSE SS ewsys}) and
the following equations
in order
to obtain the corresponding eigenvalue problem with self--energies
taken into account.

\section{Numerical method for the solution of the Bethe--Salpeter equation for QED}
\label{app. BSE QED}

The BS equation for positronium is given by
\begin{equation}\label{BSE 1 QED}
  \Gamma \left( q,P\right) =-i\frac{g^{2}}{\left( 2\pi \right) ^{4}}\int
  d^{4}k\,D^{\mu \nu }\left( q-k\right) \gamma _{\mu }G_{1}\left( k_{+}\right)
  \Gamma \left( k,P\right) G_{2}\left( k_{-}\right) \gamma _{\nu } \, .
\end{equation}
The propagators of the constituents and the corresponding
momenta are defined according to
\begin{equation}\label{BSE 1 QED 2}
  G_{i}\left( p\right) =\frac{p\hspace{-0.25cm}\diagup +m_{i}}{%
  p^{2}-m_{i}^{2}+i\epsilon }\, ,
  \quad \quad
  \begin{tabular}{l}
  $k_{+}:=k+\alpha _{1}P$ \\
  $k_{-}:=k-\alpha _{2}P$%
  \end{tabular}\, ,
  \begin{tabular}{l}
  $\alpha _{1}+\alpha _{2}=1$%
  \end{tabular}\, ,
\end{equation}
where we allowed for different masses of the electron and the positron.
Since all the calculations have been done in the Feynman gauge we used
\begin{equation}
D_{\mu \nu }\left( p\right) =\frac{g_{\mu \nu }}{p^{2}+i\epsilon }\, ,
\end{equation}
\begin{equation}
\Gamma \left( q,P\right) =\gamma _{5}\biggl( \Gamma _{1}\left( q,P\right)
+\left( P\cdot q\right) q\hspace{-0.25cm}\diagup \Gamma _{2}\left(
q,P\right) +P\!\hspace{-0.29cm}\diagup \Gamma _{3}\left( q,P\right)
 \biggr)\, ,
\end{equation}
for the propagator of the photon and for the vertex function. (For
discussion and references see section \ref{sec. BSE QED}.)
As for the scalar theory we use Gegenbauer expansions for the
scalar functions $\Gamma_{k}$
\begin{equation}
\Gamma _{k}\left( \bar{q},\bar{P}\right) =\sum_{i=0}^{\infty
}C_{i}^{1}\left( \cos \theta \right) \left[ \Gamma _{k}\right] _{i}\left(
\left\| \bar{q}\right\| ,\left\| \bar{P}\right\| \right) \, , \quad \theta
:=\sphericalangle \left( \bar{q},\bar{P}\right)  \, ,\label{X6.40}
\end{equation}
as well as for the propagators of the constituents and the photon. Applying
appropriate combinations of traces and projections one obtains a coupled
system of equations for the coefficients
$\left[ \Gamma _{k}\right] _{i}$ of the Gegenbauer
expansion (\ref{X6.40}). After Wick-rotation 
$p=(p_{0},{\bf p}) \rightarrow (ip_{4},{\bf p})=:\bar{p}$
this system of equations reads
\begin{eqnarray}
\left[ \Gamma _{1}\right] _{i} &=&\frac{g^{2}}{\left( 2\pi \right) ^{3}}%
\frac{-1}{\alpha _{1}\alpha _{2}\left( \delta +1\right) ^{2}x\eta ^{2}}\int 
\overline{d^{3}y}\,{\sf P}_{i}\sum_{i^{\prime }=0}^{\infty }C_{i^{\prime }}^{1}\left(
\cos \theta _{k}\right) S_{3}S_{2}S_{1}  \label{X11.26}
\\
&&\quad \quad \quad \quad \quad \quad \quad \quad \quad \quad \quad
\,\,\left( A_{11}\left[ \Gamma _{1}\right] _{i^{\prime }}+A_{12}\left[
\Gamma _{2}\right] _{i^{\prime }}+A_{13}\left[ \Gamma _{3}\right]
_{i^{\prime }}\right) \, , \nonumber  \\
\nonumber \\
\left[ \Gamma _{2}\right] _{i} &=&\frac{g^{2}}{\left( 2\pi \right) ^{3}}%
\frac{-1}{\alpha _{1}\alpha _{2}\left( \delta +1\right) ^{2}x\eta ^{2}}\int 
\overline{d^{3}y}\,{\sf P}_{i}\sum_{i^{\prime }=0}^{\infty }C_{i^{\prime }}^{1}\left(
\cos \theta _{k}\right) S_{3}S_{2}S_{1}\frac{1}{f_{2}} 
\\
&&\quad \quad \quad \quad \quad \left( \,\,\,\,\left[
i\left( \delta +1\right) \left( \eta x\cos \theta \right) A_{21}-\left(
\delta +1\right) ^{2}\eta ^{2}A_{31}\right] \left[ \Gamma _{1}\right]
_{i^{\prime }}\right.  \nonumber \\
&&\quad \quad \quad \quad \quad 
+\left[ i\left( \delta +1\right) \left( \eta x\cos
\theta \right) A_{22}-\left( \delta +1\right) ^{2}\eta ^{2}A_{32}\right]
\left[ \Gamma _{2}\right] _{i^{\prime }}  \nonumber \\
&&\quad \quad \quad \quad \quad
+\left. \,\,\left[ i\left( \delta +1\right) \left(
\eta x\cos \theta \right) A_{23}-\left( \delta +1\right) ^{2}\eta
^{2}A_{33}\right] \left[ \Gamma _{3}\right] _{i^{\prime }}\right) \, ,  \nonumber
\\
\nonumber \\
\left[ \Gamma _{3}\right] _{i} &=&\frac{g^{2}}{\left( 2\pi \right) ^{3}}%
\frac{-1}{\alpha _{1}\alpha _{2}\left( \delta +1\right) ^{2}x\eta ^{2}}\int 
\overline{d^{3}y}\,{\sf P}_{i}\sum_{i^{\prime }=0}^{\infty }C_{i^{\prime }}^{1}\left(
\cos \theta _{k}\right) S_{3}S_{2}S_{1}\frac{1}{f_{3}} 
\\
&&\quad \quad \quad \quad \quad
\left(
\,\,\,\,\left[ -\left( \delta +1\right) ^{2}\eta
^{2}A_{31}-x^{2}A_{21}\right] \left[ \Gamma _{1}\right] _{i^{\prime }}\right.
\nonumber \\
&&\quad \quad \quad \quad \quad
+\left[ -\left(
\delta +1\right) ^{2}\eta ^{2}A_{32}-x^{2}A_{22}\right] \left[ \Gamma
_{2}\right] _{i^{\prime }}  \nonumber \\
&&\quad \quad \quad \quad \quad
+\!\left. \left[ -\left( \delta
+1\right) ^{2}\eta ^{2}A_{33}-x^{2}A_{23}\right] \left[ \Gamma _{3}\right]
_{i^{\prime }}\,\,\,\,\right) \, ,  \nonumber
\end{eqnarray}
In addition to the dimensionless mass--parameters
\begin{equation}
\eta =\frac{M}{m_{1}+m_{2}},\quad \delta =\frac{m_{1}}{m_{2}} \, ,
\end{equation}
and the dimensionless ``momenta''
\begin{equation}
 x = \frac{\left\| \bar{q}\right\| }{m_{2}}\, ,\quad
 y = \frac{\left\| \bar{k}\right\| }{m_{2}}\, ,
\end{equation}
several abbreviations have been used:
\begin{equation}
 {\sf P}  _{i}=\frac{2}{\pi }\int_{0}^{\pi }d\theta \,\sin
^{2}(\theta)\,C_{i}^{1}\left( \cos \theta \right)
\quad \Rightarrow \quad  {\sf P} _{i} C_{k}^{1}(\cos \theta)
=\delta_{ik} \quad ,
\end{equation}
\begin{equation}
\int \overline{d^{3}y}=\int_{0}^{\infty }dy\,\int_{0}^{\pi }d\theta _{k}\sin
^{2}\theta _{k}\int_{0}^{\pi }d\varphi _{k}\sin \varphi _{k}
\quad ,
\end{equation}
\begin{eqnarray}
S_{1} &=&\sum_{n_{1}=0}^{\infty }C_{n_{1}}^{1}\left( \cos \theta _{k}\right)
\left[ g_{1}-\sqrt{g_{1}^{2}-1}\right] ^{n_{1}+1},  \label{X11.28} \\
S_{2} &=&\sum_{n_{2}=0}^{\infty }\left( -1\right)
^{n_{2}}C_{n_{2}}^{1}\left( \cos \theta _{k}\right) \left[ g_{2}-\sqrt{%
g_{2}^{2}-1}\right] ^{n_{2}+1},  \\
S_{3} &=&\sum_{n_{3}=0}^{\infty }C_{n_{3}}^{1}\left( \cos \theta _{k}\right)
\left[ g_{3}-\sqrt{g_{3}^{2}-1}\right] ^{n_{3}+1}, 
\end{eqnarray}
\begin{equation}
g_{1}=i\frac{y^{2}-\alpha _{2}^{2}\left( \delta +1\right) ^{2}\eta ^{2}+1}{%
2\alpha _{2}\left( \delta +1\right) y\eta }\,,\quad g_{2}=i\frac{%
y^{2}-\alpha _{1}^{2}\left( \delta +1\right) ^{2}\eta ^{2}+\delta ^{2}}{%
2\alpha _{1}\left( \delta +1\right) y\eta },  \label{X11.29}
\end{equation}
\begin{equation}
g_{3}=\frac{x^{2}+y^{2}}{2xy},  \label{X11.30}
\end{equation}
\begin{eqnarray}
f_{2} &=&-i\left( \delta +1\right) ^{3}\left( \eta x\cos \theta \right)
\left[ \left( \eta x\cos \theta \right) ^{2}+\eta ^{2}x^{2}\right] ,
\label{X11.31} \\
f_{3} &=&\left( \delta +1\right) ^{2}\left( -x^{2}\left( \delta +1\right)
^{2}\eta ^{2}+\left( \eta x\cos \theta \right) ^{2}\right) ,  \nonumber
\end{eqnarray}
\begin{eqnarray}
A_{11} &=&4\left( y^{2}-i\left( \alpha _{1}-\alpha _{2}\right) \left( \delta
+1\right) \left( \eta y\cos \theta _{k}\right) +\alpha _{1}\alpha _{2}\left(
\delta +1\right) ^{2}\eta ^{2}+\delta \right) ,  \label{X11.32} \\
A_{12} &=&4i\left( \delta +1\right) \left( \eta y\cos \theta _{k}\right)
\left( \left( 1-\delta \right) y^{2}-i\left( \alpha _{1}+\alpha _{2}\delta
\right) \left( \delta +1\right) \left( \eta y\cos \theta _{k}\right) \right)
,  \nonumber \\
A_{13} &=&4\left( i\left( \delta ^{2}-1\right) \left( \eta y\cos \theta
_{k}\right) -\left( \alpha _{1}+\alpha _{2}\delta \right) \left( \delta
+1\right) ^{2}\eta ^{2}\right) ,  \nonumber
\end{eqnarray}
\begin{eqnarray}
A_{21} &=&-2i\left( \delta ^{2}-1\right) \left( \eta y\cos \theta
_{k}\right) +2\left( \alpha _{1}+\alpha _{2}\delta \right) \left( \delta
+1\right) ^{2}\eta ^{2},  \label{X11.33} \nonumber \\
A_{22} &=&i\left( \delta +1\right) \left( \eta y\cos \theta _{k}\right)
\left( 2i\left( -y^{2}+\alpha _{1}\alpha _{2}\left( \delta +1\right)
^{2}\eta ^{2}-\delta \right) \left( \delta +1\right) \left( \eta y\cos
\theta _{k}\right) \right.  \nonumber \\
&&\quad \quad \quad \quad \quad \quad \quad \quad \quad \quad 
+2\left( \alpha _{2}-\alpha _{1}\right)
y^{2}\left( \delta +1\right) ^{2}\eta ^{2}  \nonumber \\
&&\quad \quad \quad \quad \quad \quad \quad \quad \quad \quad 
\left. -4i\alpha _{1}\alpha _{2}\left( \delta
+1\right) ^{3}\eta ^{2}\left( \eta y\cos \theta _{k}\right) \right) ,
\nonumber \\
A_{23} &=&-4\left( \delta +1\right) ^{2}\left( \eta y\cos \theta _{k}\right)
^{2}+2\left( \delta +1\right) ^{2}y^{2}\eta ^{2} \nonumber \\
&&\quad \quad \quad \quad \quad \quad \quad \quad \quad \quad 
+2i\left( \alpha _{1}-\alpha
_{2}\right) \left( \delta +1\right) ^{3}\eta ^{2}\left( \eta y\cos \theta
_{k}\right)  \nonumber \\
&&\quad \quad \quad \quad \quad \quad \quad \quad \quad \quad 
-2\alpha _{1}\alpha
_{2}\left( \delta +1\right) ^{4}\eta ^{4}-2\delta \left( \delta +1\right)
\eta ^{2},  \nonumber
\end{eqnarray}
\begin{eqnarray}
A_{31} &=&2\left( \delta -1\right) \left( xy\cos \omega \right) +2i\left(
\alpha _{1}+\alpha _{2}\delta \right) \left( x\eta \cos \theta \right)
\left( \delta +1\right) ,  \label{X11.34} \nonumber \\
A_{32} &=&i\left( \delta +1\right) \left( \eta y\cos \theta _{k}\right)
\left( -2\left( -y^{2}+\alpha _{1}\alpha _{2}\left( \delta +1\right)
^{2}\eta ^{2}-\delta \right) \left( xy\cos \omega \right) \right.  \nonumber
\\
&&\quad \quad \quad \quad \quad \quad \quad \quad \quad
\,\,\,\,\,\,+2i\left( \alpha _{2}-\alpha _{1}\right) \left( \delta +1\right)
y^{2}\left( x\eta \cos \theta \right)  \nonumber \\
&&\quad \quad \quad \quad \quad \quad \quad \quad \quad \quad
+\left. 4\alpha _{1}\alpha _{2}\left( \delta +1\right) ^{2}\left( \eta y\cos
\theta _{k}\right) \left( x\eta \cos \theta \right) \right) ,  \nonumber \\
A_{33} &=&-4i\left( \delta +1\right) \left( \eta y\cos \theta _{k}\right)
\left( xy\cos \omega \right) +2i\left( \delta +1\right) y^{2}\left( x\eta
\cos \theta \right)  \nonumber \\
&&\quad \quad \quad \quad \quad \quad \quad \quad \quad \quad -2\left(
\alpha _{1}-\alpha _{2}\right) \left( \delta +1\right) ^{2}\eta ^{2}\left(
xy\cos \omega \right)  \nonumber \\
&&\quad \quad \quad \quad \quad \quad \quad \quad \quad \quad
-2i\alpha _{1}\alpha _{2}\left( \delta +1\right) ^{3}\eta ^{2}\left(
x\eta \cos \theta \right)  \nonumber \\
&&\quad \quad \quad \quad \quad \quad \quad \quad \quad \quad
-2i\delta \left( \delta +1\right) \left( x\eta \cos
\theta \right) ,  \nonumber
\end{eqnarray}
\begin{equation}
   \theta     := \sphericalangle \left( \bar{q},\bar{P}\right) , \quad
   \theta_{k} := \sphericalangle \left( \bar{k},\bar{P}\right) , \quad
   \omega     := \sphericalangle \left( \bar{q},\bar{k}\right) .
   \nonumber
\end{equation}
As for the scalar BS equation we used the transformation (\ref{int.trans.})
to map the integration over the absolute value $\| \bar{k} \|$
to the range $[-1,+1]$. The integrations over $\| \bar{k} \|$ 
(corresponding to $z$ after the transf.
(\ref{int.trans.}) ) and over the angles $\theta_{k} , \varphi_{k} ,
\theta$ have been performed with the Gauss-Legendre quadrature. Finally one
obtains an eigenvalue problem for the square of the coupling constant which
can formally written as
\begin{equation}
\frac{1}{g^{2}}\left(
\begin{array}{l}
\Gamma _{1} \\
\Gamma _{2} \\
\Gamma _{3}
\end{array}
\right) =\left(
\begin{array}{lll}
B_{11} & B_{12} & B_{13} \\
B_{21} & B_{22} & B_{23} \\
B_{31} & B_{32} & B_{33}
\end{array}
\right) \left(
\begin{array}{l}
\Gamma _{1} \\
\Gamma _{2} \\
\Gamma _{3}
\end{array}
\right) .  \label{X11.36}
\end{equation}
The $B_{ik}$ are matrices of dimension
$N_{1}\left(
N_{2}+1\right) \times N_{1}\left( N_{2}+1\right) $
where $N_{1}$ is the number of mesh points that have
been used for the momentum integration over $z$ and $N_{2}$
is the maximal degree of Gegenbauer polynomials that have been
taken into account.

\section{Details of the numerical solution of the Dyson-Schwinger equation
         for the electron propagator}
\label{app. DSE QED}

Using the definition
\begin{equation}
  S\left( p\right) =\frac{1}{p\hspace{-0.25cm}\diagup A\left( p^{2}\right)
  -B\left( p^{2}\right) }
\end{equation}
of the scalar functioins $A$ and $B$ the DS equation (\ref{eq: QA DSE}) translates into
\begin{eqnarray}\label{DSE for AB}
  A\left( p^{2}\right) p\hspace{-0.25cm}\diagup +B\left( p^{2}\right)
  &=&Z_{2}\left( p\hspace{-0.25cm}\diagup -m_{0}\right) \\
  &+&iZ_{1}e^{2}\int^{\Lambda }\frac{d^{4}k}{\left( 2\pi \right) ^{4}}\left(
  \frac{-2A\left( k^{2}\right) k\hspace{-0.25cm}\diagup +4B\left( k^{2}\right)
  }{A^{2}\left( k^{2}\right) k^{2}-B^{2}\left( k^{2}\right) }\right) \frac{1}{%
  \left( p-k\right) ^{2}} \nonumber
\end{eqnarray}
Applying $\frac{1}{4}tr\left( \quad \right)$ and
$\frac{1}{4}tr\left( \gamma ^{\mu }\quad \right)$ leads to
separate but coupled equation for the scalar functions $A$ and $B$.
After Wick-rotation
$p = (p_{0},{\bf p}) \rightarrow (ip_{4},{\bf p}) =: \bar{p}$
the coupled system of equations for $A$ and $B$ reads
\begin{eqnarray}\label{DSE for AB II}
  A\left( -\bar{p}^{2}\right) =Z_{2} &+& \frac{Z_{1}e^{2}}{8\pi ^{2}}\frac{1}{%
  \bar{p}^{4}}\int_{0}^{\bar{p}^{2}}d\bar{k}\,\bar{k}^{5}\frac{A\left( -\bar{k}%
  ^{2}\right) }{A^{2}\left( -\bar{k}^{2}\right) \bar{k}^{2}+B^{2}\left( -\bar{k%
  }^{2}\right) }  \label{X12.7} \\
  &+&\frac{%
  Z_{1}e^{2}}{8\pi ^{2}}\int_{\bar{p}^{2}}^{\Lambda ^{2}}d\bar{k}\,\bar{k}%
  \frac{A\left( -\bar{k}^{2}\right) }{A^{2}\left( -\bar{k}^{2}\right) \bar{k}%
  ^{2}+B^{2}\left( -\bar{k}^{2}\right) }, \nonumber
 \end{eqnarray}
 \begin{eqnarray}
  B\left( -\bar{p}^{2}\right) =Z_{2}m_{0} &+&\frac{Z_{1}e^{2}}{2\pi ^{2}}\frac{1%
  }{\bar{p}^{2}}\int_{0}^{\bar{p}^{2}}d\bar{k}\,\bar{k}^{3}\frac{B\left( -\bar{%
  k}^{2}\right) }{A^{2}\left( -\bar{k}^{2}\right) \bar{k}^{2}+B^{2}\left( -%
  \bar{k}^{2}\right) } \label{X12.7b}  \\
  &+&\frac{%
  Z_{1}e^{2}}{2\pi ^{2}}\int_{\bar{p}^{2}}^{\Lambda ^{2}}d\bar{k}\,\bar{k}%
  \frac{B\left( -\bar{k}^{2}\right) }{A^{2}\left( -\bar{k}^{2}\right) \bar{k}%
  ^{2}+B^{2}\left( -\bar{k}^{2}\right) }\nonumber
\end{eqnarray}
where now all the momenta are euclidean.

In order to derive eqs. (\ref{X12.7}) and (\ref{X12.7b}) we used the relations
\begin{eqnarray}
  \int d\Omega _{4}\frac{1}{\left( \bar{p}-\bar{k}\right) ^{2}} &=& 2\pi
  ^{2}\left[ \Theta \left( \bar{k}^{2}-\bar{p}^{2}\right) \frac{1}{\bar{k}^{2}}
  +\Theta \left( \bar{p}^{2}-\bar{k}^{2}\right) \frac{1}{\bar{p}^{2}}\right]
  \\
  \int d\Omega _{4}\frac{\bar{p}\cdot \bar{k}}{\left( \bar{p}-\bar{k}\right)
  ^{2}}&=&\pi ^{2}\left[ \Theta \left( \bar{k}^{2}-\bar{p}^{2}\right) \frac{\bar{%
  p}^{2}}{\bar{k}^{2}}+\Theta \left( \bar{p}^{2}-\bar{k}^{2}\right) \frac{\bar{%
  k}^{2}}{\bar{p}^{2}}\right]
\end{eqnarray}
where
\begin{equation}
  \int d\Omega _{4}:=\int_{0}^{\pi }d\theta _{k}\,\sin ^{2}\theta
  _{k}\int_{0}^{\pi }d\varphi _{k}\,\sin \varphi _{k}\int_{0}^{2\pi }d\psi _{k} .
\end{equation}

The wave function renormalization constant $Z_{2}$ and the mass ratio
$m/m_{0}$ have been fixed according to
\begin{equation}
  A(0)=1 \, ,\quad B(0)=m.
\end{equation}
and the vertex renormalization $Z_{1}$ has been set equal to $1$,
for a discussion of this point see section \ref{sec. BSE QED}.

We took an exponential distribution of mesh points 
$\bar{k}_{i}$ and used an extended formula of order $O(1/N^3)$ to
perform the momentum integrations, see \cite{recipes}.

%
%

\end{document}